\documentclass{article}
\usepackage[utf8]{inputenc}
\usepackage{xcolor,soul}
\usepackage{cite}
\usepackage{graphicx}
\usepackage{booktabs}
\usepackage{array}
\usepackage{amsfonts, amsmath, amsthm, amssymb}
\usepackage{graphicx}
\usepackage{changepage}
\usepackage{color,soul}
\usepackage{framed,multirow}
\usepackage{latexsym}
\usepackage{url}
\usepackage{bm}
\usepackage{relsize}
\usepackage{hyperref}
\usepackage{siunitx}
\usepackage{bm}
\usepackage{upgreek}
\usepackage{booktabs}
\usepackage{tabularx}
\usepackage{adjustbox}

\usepackage[ruled,vlined]{algorithm2e}

\newcolumntype{L}[1]{>{\raggedright\let\newline\\\arraybackslash\hspace{0pt}}m{#1}}
\newcolumntype{C}[1]{>{\centering\let\newline\\\arraybackslash\hspace{0pt}}m{#1}}
\newcolumntype{R}[1]{>{\raggedleft\let\newline\\\arraybackslash\hspace{0pt}}m{#1}}

\usepackage{pifont}

\usepackage{tikz}

\makeatletter
\newcommand{\thickhline}{%
    \noalign {\ifnum 0=`}\fi \hrule height 1pt
    \futurelet \reserved@a \@xhline
}
\newcolumntype{"}{@{\hskip\tabcolsep\vrule width 1pt\hskip\tabcolsep}}
\makeatother

\usepackage{amssymb}
\usepackage{lipsum}
\usepackage{url}
\usepackage{enumerate}
\usepackage{amsmath}
\usepackage{array}
\usepackage{multirow}
\usepackage{booktabs}
\usepackage{soul}

\usepackage{colortbl}

\usepackage{rotating}

\usepackage{makecell} 
\graphicspath{ {./images/} }

\usepackage{geometry}
\date{ }

\title{ 
FetDTIAlign: A Deep Learning Framework for Affine and Deformable Registration of Fetal Brain dMRI}

\author{Bo Li, Qi Zeng, Simon K. Warfield, and Davood Karimi\\
\\
Department of Radiology, Boston Children's Hospital,\\ and Harvard Medical School, USA}

\begin{document}

\maketitle

\begin{abstract}

Diffusion MRI (dMRI) offers unique insights into the microstructure of fetal brain tissue in utero. Longitudinal and cross-sectional studies of fetal dMRI have the potential to reveal subtle but crucial changes associated with normal and abnormal neurodevelopment. However, these studies depend on precise spatial alignment of data across scans and subjects, which is particularly challenging in fetal imaging due to the low data quality, rapid brain development, and limited anatomical landmarks for accurate registration. Existing registration methods, primarily developed for superior-quality adult data, are not well-suited for addressing these complexities. To bridge this gap, we introduce FetDTIAlign, a deep learning approach tailored to fetal brain dMRI, enabling accurate affine and deformable registration. FetDTIAlign integrates a novel dual-encoder architecture and iterative feature-based inference, effectively minimizing the impact of noise and low resolution to achieve accurate alignment. Additionally, it strategically employs different network configurations and domain-specific image features at each registration stage, addressing the unique challenges of affine and deformable registration, enhancing both robustness and accuracy. We validated FetDTIAlign on a dataset covering gestational ages between 23 and 36 weeks, encompassing 60 white matter tracts. For all age groups, FetDTIAlign consistently showed superior anatomical correspondence and the best visual alignment in both affine and deformable registration, outperforming two classical optimization-based methods and a deep learning-based pipeline. Further validation on external data from the Developing Human Connectome Project demonstrated the generalizability of our method to data collected with different acquisition protocols. Our results show the feasibility of using deep learning for fetal brain dMRI registration, providing a more accurate and reliable alternative to classical techniques. By enabling precise cross-subject and tract-specific analyses, FetDTIAlign paves the way for new discoveries in early brain development. The code is available at \url{https://gitlab.com/blibli/fetdtialign}.

\textbf{Keywords:} Fetal brain, White matter tracts, Diffusion MRI, Registration, Spatial normalization

\end{abstract}






\section{Introduction}
\label{introduction}

Precise characterization of early brain growth in utero is crucial for understanding typical neurodevelopmental trajectories and identifying deviations that may indicate neurological disorders. This understanding is essential from a neuroscience viewpoint as the fetal period is one of the most important stages in brain development \cite{huppi1998quantitative,abe2003assessment}. Moreover, it can potentially be helpful in identifying high-risk fetuses for prenatal counseling and early developmental services. In this context, diffusion MRI offers the unique possibility to quantify the developing white matter of the fetal brain. These advanced techniques are invaluable for imaging studies that seek to understand the trajectory of brain development and the underlying neuroanatomy and neuro-microstructure \cite{mitter2015validation, Khan2018tract, Prohl2019early, wilson2021development}. Diffusion MRI measures the directional movement of water molecules within the tissue, and thus allows for the reconstruction of the brain's white matter organization. This is crucial for quantifying the connectivity and integrity of neural pathways as they emerge and mature during fetal development. By capturing subtle changes in the diffusion properties of brain tissue, these methods advance our understanding of the complex processes underlying brain development in utero.

In brain imaging studies, voxel-wise analysis that compares local image measures across groups is commonly employed, with voxel-based morphometry (VBM) being a prime example \cite{Ashburner2000vbm}. VBM offers several advantages: it is fully automated, straightforward to apply, and capable of investigating the entire brain without the need to define regions or features of interest. However, VBM relies on precise spatial correspondence across datasets, which presents a significant challenge, particularly when it is applied to diffusion imaging. Alternatively, region-of-interest and tractography-based approaches are often favored for diffusion imaging, as they operate within the space of individual subject's results, thereby avoiding the spatial correspondence issues. Nevertheless, these methods typically require user intervention to define the region to be tested. To address these limitations and combine the strengths of these approaches, tract-based spatial statistics (TBSS) was proposed \cite{Smith2006tract}.

TBSS, as implemented in the FMRIB Software Library (FSL) \cite{Smith2004fsl}, was introduced to mitigate the effects of residual misalignment during the deformable registration of diffusion imaging. This approach facilitates localized cross-subject statistical analysis and eliminates the need to set a kernel size for smoothing as required in VBM \cite{Jones2005effect}. Additionally, TBSS results are more reproducible than those of VBM or techniques that rely on manual ROI placement \cite{Smith2014chapter}. However, TBSS still requires that images be aligned in a common space, which presents significant challenges. Establishing anatomical correspondence across subjects is particularly difficult for low-quality data or for subjects with rapidly changing imaging characteristics, such as fetal brain images, which is the focus of this study. While the skeletonization and projection procedures in TBSS may compensate for minor misalignments, they can disrupt topological consistency when misalignments are substantial. Since only voxels containing locally maximal fractional anisotropy (FA) values are independently projected onto the skeleton without considering fiber tract geometry, regions between two skeleton points can be artificially split across multiple anatomical locations \cite{Zalesky2011Moderating}. Furthermore, the construction of the skeleton across the entire white matter does not differentiate between adjacent white matter tracts. Consequently, voxels with locally maximal FA values may be misassigned to neighboring fiber tracts within the projection's search range, such as the inferior longitudinal and inferior frontal-occipital fasciculi \cite{Bach2014methodological}.

To enhance the original TBSS pipeline, prior efforts have focused on its key components: 1) improving registration quality to achieve better anatomical correspondence across subjects, 2) refining the representation of tract-specific diffusion measures to increase specificity and reduce the influence of adjacent tracts \cite{Gong2005Asymmetry, Corouge2006Fiber, ODonnell2009Tract, Yushkevich2009Structure, Pecheva2017TSA}, and 3) advancing the statistical analysis of the derived measures.

Image registration is a powerful tool in medical imaging. It estimates a displacement field that describes how points in one image should move to optimally match the anatomically corresponding points in another image so that the same structures or image features can be compared across scans or subjects. It is however a complex and ill-posed estimation problem and presents additional challenges when co-registering matrix-encoded dMRI data compared with scalar images, such as T1-weighted MRI \cite{ruiz2002nonrigid}. The spatial registration of dMRI has been performed using various diffusion-based measures, including FA scalar maps derived from diffusion tensor imaging (DTI) \cite{ball2010optimised, deGroot2013improving, Schwarz2014Improved, wang2014diffusion, Li2021Segis}, diffusion tensors \cite{alexander2000elastic, alexander2001spatial, Zhang2007High, Khan2018tract, Irfanoglu2016DRTAMAS, commowick2008continuous,zhang2007unbiased}, orientation distribution functions \cite{Raffelt2011Symmetric,Zhang2021ddmreg,Commowick2015,Karimi2023tbss}, tractography streamlines \cite{Garyfallidis2015robust, garyfallidis2018recognition,o2012unbiased}, and multi tensor-based models \cite{Taquet2013mathematical, Jbabdi2010Crossing}. 

Several studies have explored updating the TBSS pipeline with contemporary advancements in image registration techniques. For instance, \cite{deGroot2013improving} evaluated the feasibility of replacing TBSS's two-step registration-projection approach with a single registration step, finding that better alignment could be achieved by upsampling diffusion tensor rather than upsampling tensor-derived FA images when bringing the FA maps to the standard MNI152 space. \cite{Bach2014methodological} suggested using a high-resolution DTI template as the common space target for registration instead of using the default FA template. Similarly, improvements were observed when group-wise registration was employed to spatially normalize FA images to a study-specific template, \cite{Keihaninejad2012importance, Li2021LUGR}. Improved registration increases the longitudinal test-retest reliability of TBSS \cite{madhyastha2014longitudinal} and allows for voxel-wise analysis that outperforms TBSS \cite{Schwarz2014Improved}.

\begin{figure*}[ht]
\caption{Illustrated summary of the study.}
\centering
\includegraphics[width=1.0\textwidth]{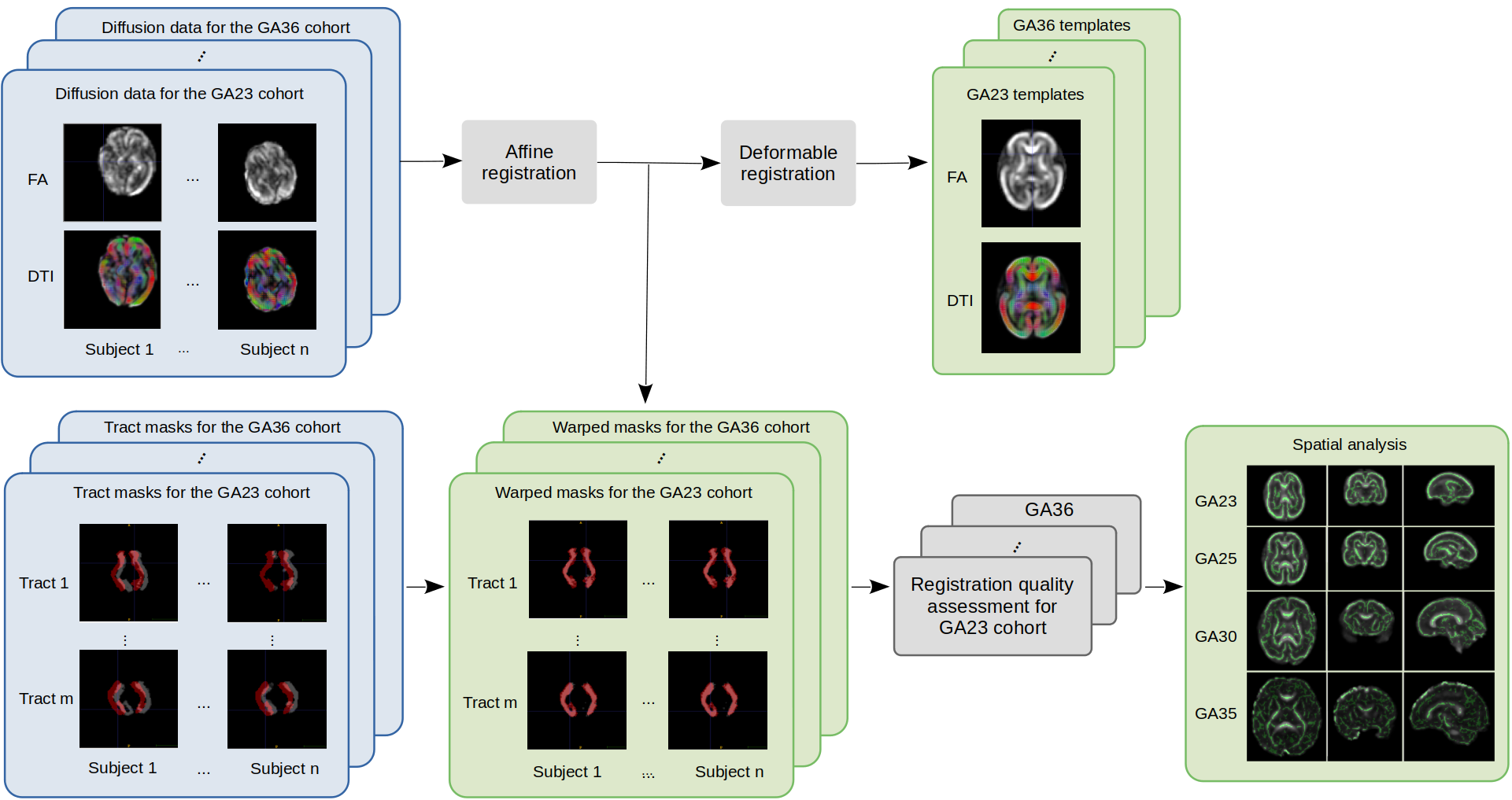}
\label{fig:overall_framework}
\end{figure*}

Nevertheless, achieving accurate in utero assessments of fetal brain development is fraught with complexities due to the dynamic and rapidly developing nature of the fetal brain during early growth, the small size of the developing structures, fetal and maternal motion, and the inherent limitations in resolution and signal-to-noise ratio of current imaging techniques. These complexities have precluded the use of TBSS for studying brain development in the human fetus. In this study, we lay the foundation by introducing the first computational method designed for accurate spatial alignment of fetal brain dMRI, i.e., FetDTIAlign, enabling voxel-wise and tract-specific imaging studies of the human fetus. Our method leverages advances in learning-based techniques for both affine and deformable registration, as well as recent availability of high-quality spatiotemporal fetal DTI atlases. We anticipate that FetDTIAlign, which integrates tract segmentation with a DTI-based high-dimensional registration framework, will ensure precise focal and geometric correspondence across fetal brains, enabling more accurate modeling of the early stage of brain's developmental trajectory. Figure~\ref{fig:overall_framework} provides a graphical overview of this study.

\section{Background}

\subsection{Diffusion Tensor Imaging (DTI)}\label{ss:DTI}

DTI is one of the local diffusion models that characterizes the diffusion of water molecules in biological tissues, providing insight into the microstructural integrity of white matter in the brain \cite{pierpaoli1996toward}. The core of DTI is the diffusion tensor, a mathematical construct represented as a $3 \times 3$ symmetric positive-definite matrix, $\mathbf{D}$, which describes the diffusion process in three-dimensional space: 
$\mathbf{D} = 
\begin{pmatrix}
D_{xx} & D_{xy} & D_{xz} \\
D_{xy} & D_{yy} & D_{yz} \\
D_{xz} & D_{yz} & D_{zz}
\end{pmatrix}$.
This tensor is estimated from diffusion MRI acquired in multiple gradient directions \cite{basser1994estimation}. Mathematically, the diffusion-weighted signal $S(\mathbf{g})$ in the presence of a diffusion gradient $\mathbf{g}$ can be expressed as:
\begin{equation}
    S(\mathbf{g}) = S_0 \exp\left(-b \mathbf{g}^\top \mathbf{D} \mathbf{g}\right),
\end{equation}
where:
$S_0$ is the signal intensity without diffusion weighting,
$b$ is the diffusion weighting factor (i.e., b-value), and
$\mathbf{g}$ is the unit vector in the direction of the applied gradient (i.e., b-vector).

To derive bio-physically meaningful measures from the diffusion tensor, the eigenvalues ($\lambda_1$, $\lambda_2$, $\lambda_3$) of the tensor are calculated. These eigenvalues correspond to the principal diffusivities along the major, medium, and minor axes of diffusion. From these eigenvalues, several scalar diffusion measures are computed, each providing different insights into tissue microstructure. Fractional anisotropy (FA) quantifies the degree of anisotropy of diffusion, indicating how directional the diffusion is within a voxel:
\begin{equation}
    \text{FA} = \sqrt{\frac{3}{2}} \cdot \frac{\sqrt{(\lambda_1 - \bar{\lambda})^2 + (\lambda_2 - \bar{\lambda})^2 + (\lambda_3 - \bar{\lambda})^2}}{\sqrt{\lambda_1^2 + \lambda_2^2 + \lambda_3^2}},
\end{equation}
where $\bar{\lambda} = \frac{\lambda_1 + \lambda_2 + \lambda_3}{3}$ is the mean diffusivity (MD).

These diffusion measures are fundamental in analyzing the microstructural properties of brain tissues and form the basis for more complex analyses, which will be described in the following sections.

\subsection{Tract-Based Spatial Statistics (TBSS)}\label{ss:background_tbss}

The standard TBSS pipeline \cite{Smith2006tract} primarily concerns the FA maps computed from the DTI of a given group of subjects. It starts by registering all subjects’ FA images into a common space by linear and nonlinear registration, using FLIRT and FNIRT approach as available in FSL \cite{andersson2007non, greve2009accurate}. Three choices of the \textit{common space template} (i.e., target image) are provided: (1) an existing image within the group that is identified as the most typical subject in the group, who has a minimum mean deformation to all other subjects, (2) a standard template image, for instance, ``FMRIB58\_FA\_1mm" in the MNI152 space, and (3) a user-specified template. Once all subjects’ FA images are aligned to the chosen target, the transformed images are averaged to create a \textit{mean FA image}. Subsequently, an \textit{FA skeleton} is computed from the mean FA image, the local surface perpendicular to the skeleton sheet is computed, and a search is performed in this direction to identify the voxel with the highest FA, which is deemed to represent the center of the tract. This is further filtered by an FA threshold in order to restrict further analysis to only the voxels that are within white matter and have good correspondence across subjects. Finally, each subject’s aligned FA image is “projected” onto the mean FA skeleton. At each point in the skeleton, a given subject’s FA image is searched in the perpendicular tract direction to ﬁnd the maximum FA value, and assign this value to the skeleton voxel. This results in a 4D volume of \textit{all FA skeletonized}. The aim here is to account for residual misalignment between subjects after the nonlinear registrations by achieving alignment between the skeleton and the subject’s FA image, without requiring perfect nonlinear registration. However, the misalignment in the direction parallel to the tract is not considered \cite{Bach2014methodological}. 

\subsection{Tensor reorientation}\label{ss:background_reorientation}

Spatial registration of orientation-encoded vector images, such as diffusion tensors and diffusion-weighted volumes, requires not only anatomical correspondence at the voxel level but also directional correspondence of underlying white matter structures. This necessitates additional reorientation operation for each voxel to ensure alignment of the directional information, which essentially serves to correct for the cumulative rotational effects caused by the series of transformations applied to the corresponding tensors. The key step in reorienting a tensor is, therefore, to accurately estimate the overall rotation occurring during the transformation of each voxel. 

In the context of a linear transformation, consider a square \( n \times n \) invertible real matrix $A$ that defines a linear (non-translational) transformation on $\mathbb{R}^n$, mapping a column vector $x$ to $Ax$. The polar decomposition of $A$ expresses it as the product $A=RP$, where the factor $R$ is an \( n \times n \) real orthogonal matrix, and the factor $P$ represents a positive-definite matrix \cite{Hall2015Lie}. In this decomposition, $R$ represents a rotation (or reflection) in $\mathbb{R}^n$, while $P$ represents a scaling along the $n$ orthogonal axes. Thus, the polar decomposition can be interpreted as separating the linear transformation defined by $A$ into two distinct components: a scaling of the space along orthogonal directions ($P$), followed by a rotation ($R$) in $\mathbb{R}^n $. 

The closest rotation matrix to $A$ can be computed in several ways. One approach is to use Gram-Schmidt orthogonalization. Alternatively, the polar decomposition can be reformulated by computing $S=\sqrt{A^T A}$ and then obtaining \( R \) as $R=AS^{-1}$. Another method involves minimizing the Frobenius norm, which measures the Euclidean distance between $A$ and $R$, subject to the constraint $det(R)=+1$ \cite{mitter2015validation}. Additionally, the rotation can be parameterized explicitly and incorporated within the objective function \cite{zhang2006deformable}. In practice, when $P$ is positive-definite and $R$ is orthogonal, the existence of the singular value decomposition (SVD) of $A$ (i.e., $A=W \Sigma V^T$) is equivalent to the existence of polar decomposition. Consequently, \( R \) can be obtained directly from the SVD of $A$ by setting $R=WV^T$ and $P=V \Sigma V^T$.

For a non-linear deformation $\phi$, if $\phi$ is differentiable at a location $x$ in $\mathbb{R}^n$, then its differential at that location is represented by the Jacobian matrix $J\phi(x)$. The linear transformation given by $J\phi(x)$ provides the best linear approximation of $\phi$ in the vicinity of $x$. The rotation component of $\phi$ at $x$, i.e., $R_x$, can then be extracted through the polar decomposition of the Jacobian matrix $J\phi(x)$.

\section{Methods}

Given a group of $n$ diffusion measure maps $\mathcal{I} = \{I_1,\, ..., I_n \}$, each described by scalar or vector values $I_i(x)$ at spatial coordinate $x \in X_i , i=1...n$, the spatial normalization procedure estimates a set of transformations $\tau = \{\boldsymbol{\mathcal{T}}_1, ..., \boldsymbol{\mathcal{T}}_n\}$ that aligns the images $\mathcal{I}$ to a common reference space $\overline X$, such that the anatomically corresponding structures are of the same coordinate, and the underlying diffusion tensors are of the same orientation, i.e., $I_i(\boldsymbol{\mathcal{T}}_i) \approx I_j(\boldsymbol{\mathcal{T}}_j), \, \forall i\neq j$.

In this study, we employed a set of high-quality spatiotemporal fetal brain DTI template between 23 and 36 weeks of gestation \cite{calixto2023detailed} as the common reference for spatial normalization. For each GA group per week, denoted as $\mathcal{I}_{GA}$, we registered all individual images in $\mathcal{I}_{GA}$ to the corresponding GA-specific DTI and FA template $\mathcal{A}_{GA}$. Thereafter, it allows for various spatiotemporal imaging studies, including discovering both tract-specific and voxel-wise imaging difference across groups, and modeling the early trajectory of fetal brain development. To serve as an end-to-end learning-based alternative to classical registration algorithms, we performed hierarchical optimization scheme to estimate both an affine $\phi_A$ and a residual nonlinear deformation $\phi_D$, and performed the tensor reorientation during optimization. 

Lastly, to minimize the smoothing effects that can result from multiple interpolation steps, we applied the composed deformation field \( \phi = \phi_A \circ \phi_D \) directly to the original images in a single interpolation operation.

\subsection{Affine registration}
\begin{figure*}[!h]
\caption{Affine registration approach in FetDTIAlign. The fetus shown in this figure has a gestational age of 25 weeks.}
\centering
\includegraphics[width=1.0\textwidth]{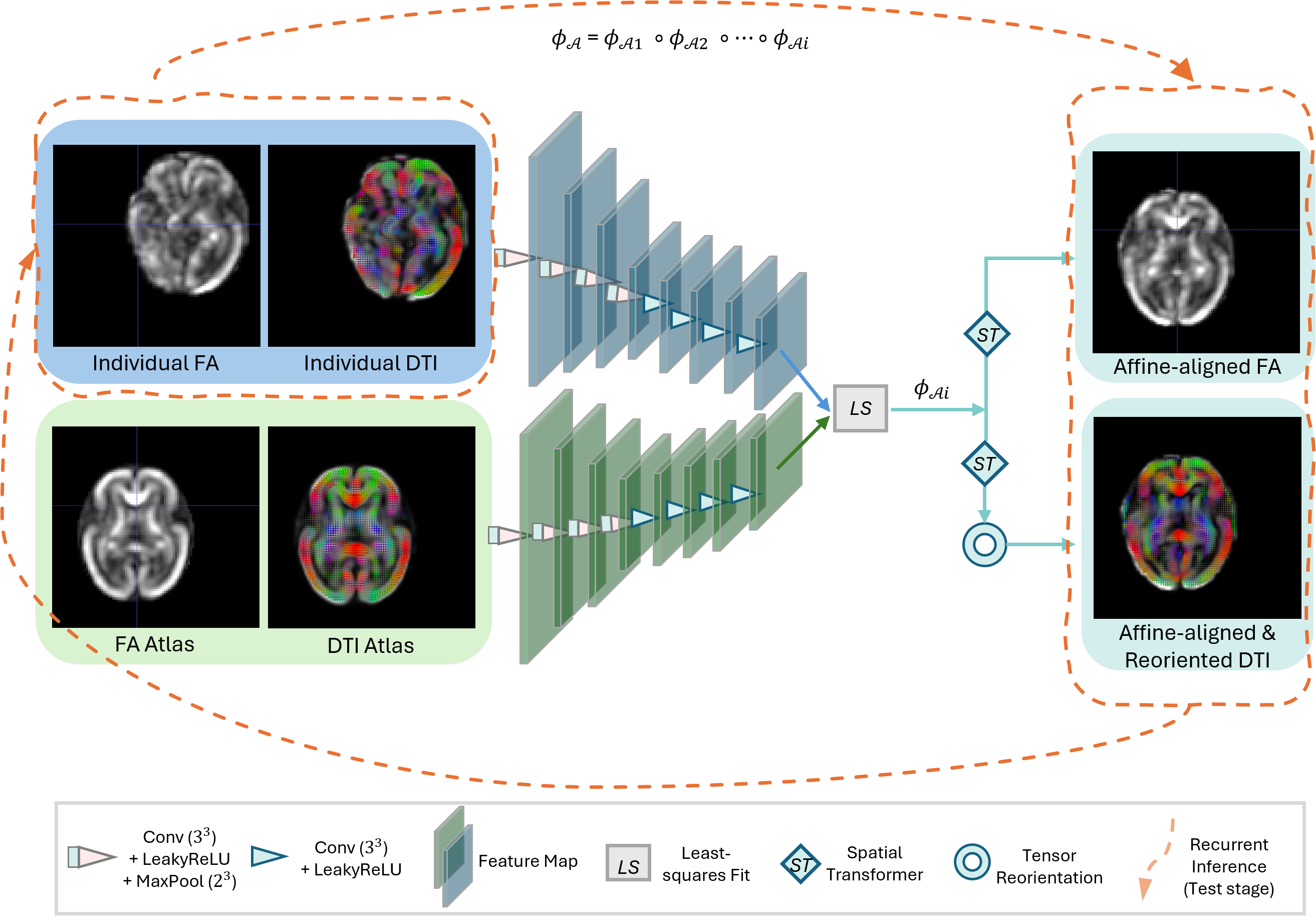}
\label{fig:affine_framework}
\end{figure*}

Affine transformation between the fetal brain and the age-matched template was estimated from the feature embedding of the concatenated FA and diffusion tensor images (Figure~\ref{fig:affine_framework}). To improve registration robustness against varying initial positions, we applied extensive yet practical data augmentation to the training samples for both individual and template images. For each augmented transformation, we uniformly sampled the following random parameters: an isotropic scaling factor $s \in [0.9, 1.1]$, a rotation angle $\theta_x, \theta_y, \theta_z \in [-20^\circ, 20^\circ]$ around the image center and along only one of the x, y, and z axes, and a translation vector $t_x, t_y, t_z \in [-4, 4]$ voxels for translation along the respective axes. To ensure consistent data formatting and tensor reorientation, we followed the same procedure employed for template construction \cite{zhang2006deformable} to warp the FA and diffusion tensor images using the synthesized affine matrices. 

The transformation was obtained through a learning-inference process, where an affine matrix $\hat{\phi}_A$ was estimated from a trainable model $G_\Psi$, with the trainable parameters $\Psi$ optimized to minimize the dissimilarity between the registered subject and template. Specifically, for any image $I_i$ in a given gestational age group, the affine matrix between the image and the corresponding template $\mathcal{A}_{GA}$ was estimated as follows:

\[
\hat{\phi}_{A} = G_\Psi(\mathcal{A}_{GA}, I_i), 
\]
and the performance metric was computed to drive the update of $\Psi$. We employed a multi-task objective function $\mathcal{L}_{\phi_A}$ to assess regional structural dissimilarity ($\mathcal{L}_{FA}$) of warped FA images using local normalized cross-correlation with a cubic kernel of size $9^3$ voxels, and a voxel-wise tensor dissimilarity ($\mathcal{L}_{DTI}$). Normalized cross-correlation is a widely used metric for assessing structural alignment in medical images owing to its robustness against intensity variations \cite{klein2009elastix}. For quantifying the distance between pairs of diffusion tensors, various metrics have been proposed in the literature \cite{alexander2000elastic,zhang2006deformable}. In this work, we used a combined metric of the following form:

\begin{align}
\mathcal{L}_{DTI} =  
\frac{1}{|\Omega_\mathcal{A}|} \sum_{\Omega_\mathcal{A}} \Big( & \text{ED}(\mathcal{A}_{GA}, R(I_i \circ \hat{\phi}_{A})R^T) \nonumber \\
& + \text{DD}(\mathcal{A}_{GA}, R(I_i \circ \hat{\phi}_{A})R^T) \Big).
\end{align}

In this equation, $R$ is the overall rotation component decomposed (Section~\ref{ss:background_reorientation}) from the estimated affine transformation $\hat{\phi}_{A}$ and applied to the $3 \times 3$ diffusion tensor matrix (Section~\ref{ss:reorientation}) at each voxel. We employed a combination of the Euclidean Distance (ED) of the full tensor and the Diagonal Frobenius Distance (DD) to optimize the alignment. This hybrid approach provides a comprehensive measure of diffusion dissimilarity on both the full tensor’s geometric information and specifically the diffusion along the primary anatomical axes while keeping the method computationally efficient. The metrics were averaged over the reoriented tensors $\mathbf{D}_{i,\mathcal{A}}$ within the brain region of the template $\Omega_\mathcal{A}$:

\begin{equation}
\text{ED} (\mathbf{D}_\mathcal{A},\mathbf{D}_{i,\mathcal{A}}) = \|  \mathbf{D}_\mathcal{A} - \mathbf{D}_{i,\mathcal{A}}   \|^2_2 ,
\end{equation}

\begin{align}
\text{DD} (\mathbf{D}_\mathcal{A}, \mathbf{D}_{i,\mathcal{A}}) = &(D_{\mathcal{A}: xx} - D_{i,\mathcal{A}: xx})^2 +  (D_{\mathcal{A}: yy} - D_{i,\mathcal{A}: yy})^2  \notag \\
& +  (D_{\mathcal{A}: zz} - D_{i,\mathcal{A}: zz})^2 .
\end{align}
Combining the two similarity terms, $\mathcal{L}_{\phi_A}$ was defined as $\lambda \mathcal{L}_{FA}$ + $\mathcal{L}_{DTI}$, where $\lambda$ was empirically optimized for better validation performance.

The affine registration model consisted of two domain-specific encoders to generate feature embeddings, a registration head that estimates an optimal affine transformation matrix between the embeddings, and a tensor reorientation layer designed to correct for rotational effects on the underlying diffusion tensor orientations (Figure~\ref{fig:affine_framework}). We implemented a parallel-encoder architecture with two identical feature encoders, each consisting of four convolutional layers followed by LeakyReLU activations \cite{maas2013rectifier} and max-pooling (down-sampling factor of 2) \cite{lecun1998gradient}, along with four additional convolutional layers without down-sampling. Since fully convolutional networks are not inherently rotation- or translation-invariant, we used separate weights for the affine registration encoders to allow each to specialize in the moving and target domains, better handling the large positional variance before alignment. The features, shaped as $8^3 \times 32$ at the most abstract level for both the moving and target domains, were aggregated to estimate the twelve affine parameters by aligning the mass centers of the feature embeddings using a least-squares fit \cite{hoffmann2024anatomy, wang2024brainmorph,moon2000mathematical}. To ensure sufficient capacity for handling large global transformations, we applied recurrent inference during testing, running the model $i$ times. Each subsequent inference used the warped output of the previous inference as input. In this unsupervised setting, where target images were available, we computed the mean squared error (MSE) of the registered FA and diffusion tensor images after each inference. The final result was chosen based on the inference that yielded the lowest MSE for both FA and DTI. Specifically, if the lowest MSE occurred at the $i$th inference, the final affine transformation was obtained by composing the accumulated transformations from all previous steps: \( \hat{\phi}_{A} = \hat{\phi}_{A1} \circ \hat{\phi}_{A2} \circ ...\circ \hat{\phi}_{Ai} \). This transformation was then applied to warp the original moving image.

\subsection{Deformable registration}

\begin{figure}[!h]
\caption{Deformable registration approach in FetDTIAlign using an FA encoder (in cyan), a Tensor encoder (in purple), and a multi-scale recurrent registration decoder (in grey). The fetus shown in this figure has a gestational age of 25 weeks.}
\centering
\includegraphics[width=1.0\textwidth]{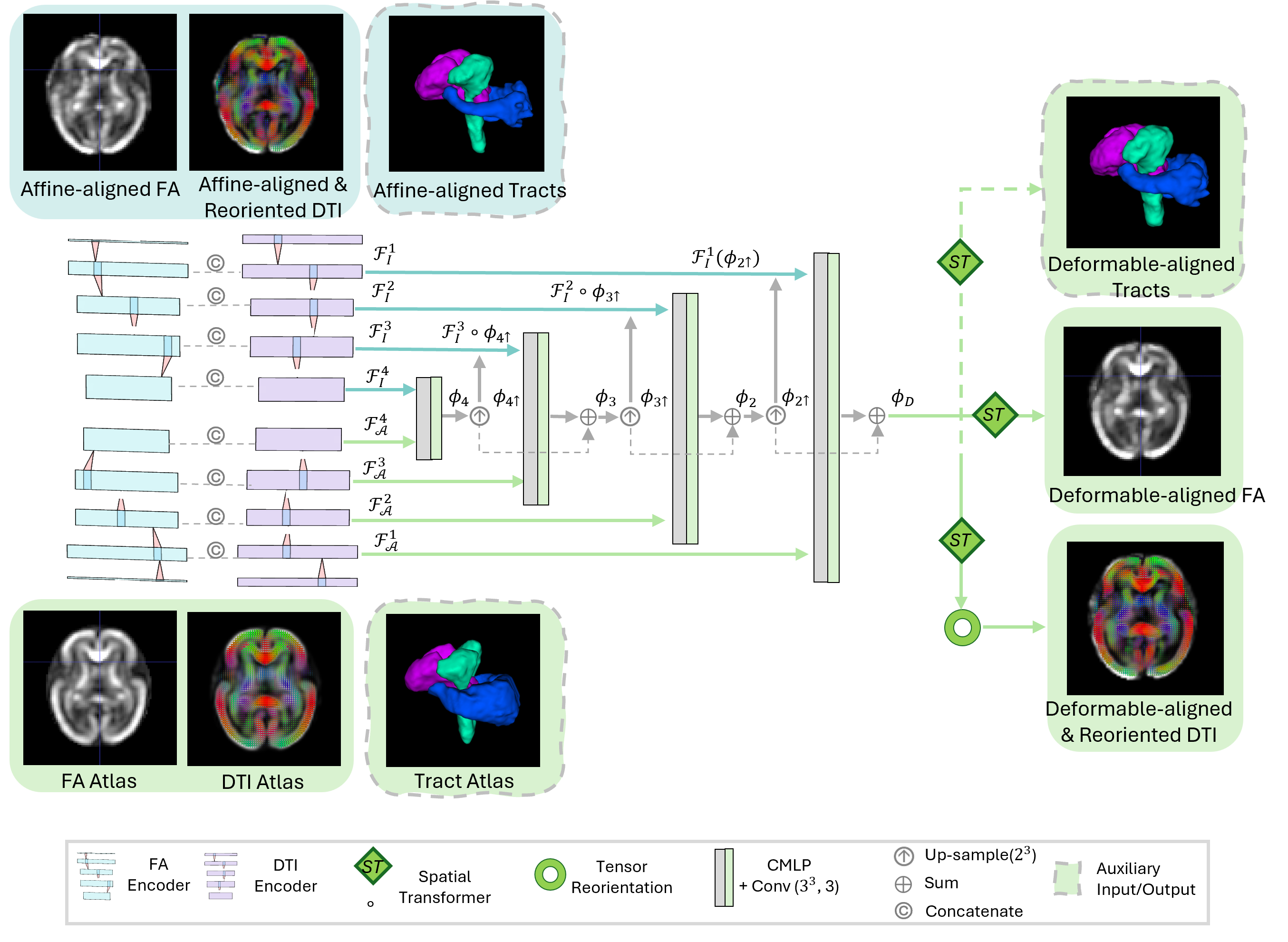}
\label{fig:deformable_framework}
\end{figure}

Deformable registration between the affine-aligned fetal brain and the corresponding template was estimated using the FA and diffusion tensor images. During model training, tract segmentation masks (Section~\ref{ss:segmentation}) were optionally incorporated in the loss function when available. Unlike the affine registration stage, where both images were concatenated and input as a whole, in the deformable registration stage, we extracted coarse-to-fine feature embeddings separately from scalar FA images ($\mathcal{F}_I^{j,FA}$ , $\mathcal{F}_\mathcal{A}^{j,FA}$ , $j=1...k$) and vector-encoded diffusion tensor images ($\mathcal{F}_I^{j,DTI}$ , $\mathcal{F}_\mathcal{A}^{j,DTI}$,  $j=1...k$) using an FA Encoder and a Tensor Encoder with $k$ hierarchical scales (Figure~\ref{fig:deformable_framework}). The use of `modality'-specific encoders allowed us to derive complementary features from each, which were then concatenated at each scale ($\mathcal{F}_I^j$, $\mathcal{F}_\mathcal{A}^j$ , $j=1...k$) for recurrent dense deformation estimation through a registration decoder. The entire process can be formulated as: 
\[
\hat{\phi}_{D}^j = H_\Theta(\mathcal{F}_\mathcal{A}^j, \mathcal{F}_I^j \circ \hat{\phi}_{D}^{j-1}). 
\]

For deformable registration, both the FA and Tensor encoders utilized a symmetric Siamese architecture with shared weights for the affine-aligned moving and template images. After affine registration, the relative positions and orientations of the images were well aligned, reducing the impact of the inherent lack of rotation and translation invariance in convolutional networks. Consequently, shared weights between the encoders can be employed at this stage, as the images contain similar spatial information, enabling more effective generalization and minimizing the risk of overfitting. Each encoder operated at four different scales (i.e., $k=4$), with each scale composed of two convolutional layers followed by average pooling with a down-sampling factor of 2. 

For the decoder, we adopted the classical multi-resolution scheme, where the deformation field estimated in a coarser level was up-sampled and linearly interpolated to provide an initial deformation in the next finer level \cite{ruiz2002nonrigid,shen2002hammer}. To estimate the residual deformation at each resolution level, we used warped feature embeddings ($\mathcal{F}_I^j \circ \hat{\phi}_{D}^{j-1}$) instead of warping the original moving images. This approach was expected to preserve the sparse details in fetal brain diffusion images by minimizing further interpolation and smoothing. Additionally, it eliminated the need for tensor reorientation at each resolution and leveraged the high-dimensional features, which are potentially more discriminative for image registration. For each resolution level, to optimize structural similarity across fine-grained local regions and broader spatial neighborhoods between warped feature embeddings, we employed correlation-aware multi-window multi-layer perceptron (CMLP) blocks \cite{meng2024correlation}. These blocks compute local correlations and capture multi-range dependencies. Specifically, each block leverages multiple receptive field sizes, using window sizes of \{$3^3, 5^3, 7^3$\}, to aggregate spatially varying contextual information, which is integrated into the deformation field estimation. This multi-scale strategy allows for precise local alignment while ensuring spatial coherence over a broader context, facilitating the registration of regions with complex diffusion orientations.

We optimized the trainable parameters using a multi-task objective function with the finest scale deformation, $\hat{\phi}_{D}$. Similar to the affine stage, our objective function measured structural FA similarity $\mathcal{L}_{FA}$ (with a kernel of size $5^3$ voxels) and voxel-wise tensor similarity $\mathcal{L}_{DTI}$. Additionally, the objective function included spatial correspondence of white matter tracts $\mathcal{L}_{tract}$ quantified with the multi-class average Dice coefficient between aligned tract masks within the brain region of the tract atlas $\Omega_\mathcal{A}$. Meanwhile, to penalize large deviations of deformation and preserve anatomical topology during transformations, a deformation smoothness term $\mathcal{L}_{def}$ was included, which computes the average spatial gradients of the deformation field over all voxels,
\begin{equation}\label{eq:Ldef}
\mathcal{L}_{def}(\hat{\phi}_D) = \frac{1}{|\Omega|} \sum_{x \in \Omega} \| \nabla \hat{\phi}_D (x) \|^2_2.
\end{equation}
Combining the terms, $\mathcal{L}_{\phi_D}$ was defined as $\lambda \mathcal{L}_{FA}$ + $\mathcal{L}_{DTI}$ + $\mathcal{L}_{tract}$ + $ \gamma \mathcal{L}_{def}$, where $\lambda$ and $\gamma$ were empirically optimized to achieve better validation performance.

\subsection{Tensor reorientation}\label{ss:reorientation}

Reorientation of the diffusion tensor was performed in order to ensure that the local orientation of diffusion tensors remains consistent with the anatomy after an image transformation. We implemented tensor reorientation as a general computation layer within the trainable models for both affine and deformable registration. Each estimated affine matrix, as well as the residual nonlinear deformation, can be considered as a dense deformation field that describes the new position $x'$ of each point after transformation. The differential of this field, represented by the Jacobian matrix $J\phi(x)$, varies across locations and captures the local linear transformations, including rotation, scaling, and shearing. To isolate the rotation component $R$, we applied SVD to the Jacobian matrix, separating it from the scaling component, as explained in Section~\ref{ss:background_reorientation}:
\[
J\phi(x) = \frac{\partial x'}{\partial x},
\]

\[
\text{SVD}: J\phi(x) = W \Sigma V^T,
\]

\[
R = WV^T.
\]

\section{Experiments}
To evaluate the proposed FetDTIAlign, we applied it to spatially normalize a fetal brain MRI dataset to a template spanning 23 to 36 weeks of gestation (Section~\ref{ss:datasets}). We assessed the quality of both affine and deformable registration (Section~\ref{ss:metrics}) and compared the results with state-of-the-art classical and learning-based algorithms (Section~\ref{ss:baseline_methods}).

\subsection{Study population and data acquisition}\label{ss:datasets}
In-utero fetal MRI scans analyzed in this study were acquired as part of a prospective research study at Boston Children’s Hospital (BCH) in Boston, MA. All participants provided written informed consent, and the BCH Institutional Review Board (IRB) approved the study while following HIPAA guidelines. The inclusion criteria for the study were mothers between 18 and 45 years of age who had normal pregnancies with gestational ages between 22 and 37 weeks. Exclusion criteria included any contraindications to MRI, high-risk pregnancies, fetal central nervous system anomalies, and maternal comorbidities such as diabetes, hypertension, or substance abuse. Imaging data were acquired using 3T MRI scanners. For diffusion MRI scans, 2-8 echo-planar diffusion-weighted volumes were acquired along orthogonal planes relative to the fetal head. Each acquisition included b-values of 0 s/$mm^2$ (1-2 volumes), and 500 s/$mm^2$ (12-24 volumes), TR of 3000-4000 ms, TE of 60 ms, in-plane resolution of 2 mm, and slice thickness of 2-4 mm.

\begin{figure}[!h]
\caption{The number of subjects in each Gestational Age (GA) group by Training and Test split for the BCH dataset}. Overlapping subjects exist across age groups due to the inclusion criteria (Section~\ref{ss:datasets}). However, each image pair, whether for training or testing, is unique. For quantitative evaluation, the test age groups consists of completely non-overlapping subjects to ensure valid reporting.
\centering
\includegraphics[width=0.6\textwidth]{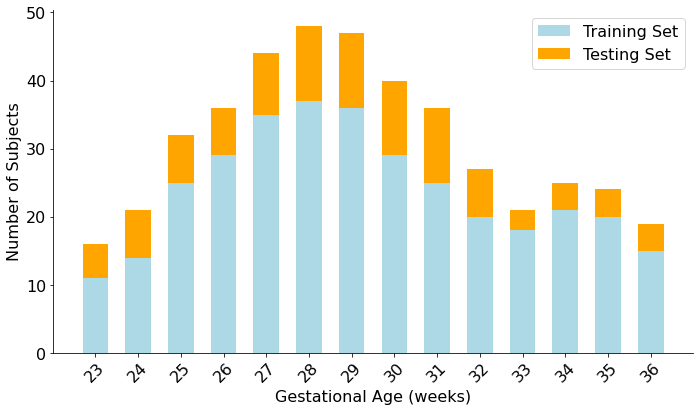}
\label{fig:training_test_split}
\end{figure}

The DTI templates were computed at one-week intervals using data from fetuses with gestational ages within the range of $[GA-1, GA+1]$ weeks. We conducted quantitative evaluations on data from three gestational age groups—GA25, GA30, and GA35—representing different stages of brain development and different tissue contrast. These groups consisted of diffusion data from 7, 11, and 4 subjects, respectively. The data used in the validation phase remained unseen for the learning-based methods during model development phase. For the training of the learning-based methods, we expanded the range to include subjects with gestational ages within $[GA-2, GA+2]$ weeks for each group, increasing the variation in the image pairs used for registration. This resulted in a training set of 335 image-atlas pairs, and a clean test-set of 101 pairs. The subject-wise data split is shown in Figure~\ref{fig:training_test_split}. In addition, to investigate the optimal configuration of classical algorithms for this dataset, we conducted qualitative evaluations on data from the youngest age group, GA23, which represents a challenging scenario for approaches developed for adult image analysis.

In addition to the in-house fetal MRI scans obtained at BCH, we included data from $n=35$ fetuses from the Developing Human Connectome Project\footnote{dHCP: \url{https://www.developingconnectome.org/}} for external testing. This dataset provides research-grade quality dMRI data acquired using a multi-shell scheme with b-values of 0 s/$mm^2$ (15 volumes), 400 s/$mm^2$ (46 volumes), and 1000 s/$mm^2$ (80 volumes), totaling 141 volumes \cite{christiaens2019fetal}. To investigate the generalizability of the proposed method for higher b-value images, for this test dataset we used the diffusion-sensitized volumes at b=1000 s/$mm^2$ to compute DTI. To ensure a well-distributed evaluation across different developmental stages, we randomly sampled data from fetuses with gestational ages spanning from 22 to 37 weeks, included at least two fetuses from each gestational week.

\subsection{Image preprocessing and white matter tract segmentation}\label{ss:segmentation}

Diffusion MRI data from BCH were pre-processed using a previously validated pipeline \cite{marami2016motion}. This pipeline used a Kalman filtering-based algorithm to track the fetal head motion and performs slice-to-volume registration to align the q-space dMRI data in a standard atlas space. The reconstructed q-space volume had an isotropic voxel size of 1.2 mm. 

White matter tract segmentation masks for fetuses included in the BCH data were obtained in a prior work \cite{calixto2025detailed} using a well-validated anatomically constrained fetal tractography method \cite{calixto2024anatomically}. In brief, after computing a whole-brain tractogram for each fetus, an automatic tool \cite{wassermann2016white} was used to extract individual streamline bundles representing well-defined tracts. Experts with extensive knowledge of fetal neuroanatomy and neuroradiology verified the tract extractions. Subjects with inferior tract extraction results were excluded. In order to convert the streamline bundles to binary tract masks for use in this study, we computed for each tract a streamline density map and rejected voxels containing less than 5 percentile of streamlines for that tract. A total of 60 white matter tracts were extracted for each fetus, which were reduced to 34 tract labels after combining the bilateral tracts. Tract names are presented in Table~\ref{tab:summary_tracts}. Further anatomical description of the tracts can be found in \cite{calixto2025detailed}.

\begin{table}[]
    \centering
    \scriptsize
    \begin{tabular}{c|c}
        \hline
        \textbf{Category} & \textbf{Details} \\ 
        \hline
        \multirow{4}{*}{\textit{Commissural tracts}} & \makecell{Corpus Callosum (CC): \\ the Rostrum part (CC 1), Genu part (CC 2), Rostral body (CC 3), \\ Anterior midbody (CC 4),  Posterior midbody (CC 5), Isthmus part (CC 6) \\ and Splenium part (CC 7)} \\
        \hline
        
        \multirow{12}{*}{\textit{Projection tracts}} & \makecell{Corticopontine-cerebellar fibers: \\ Fronto-pontine tract (FPT), and Parieto-occipital pontine (POPT) } \\
        \cline{2-2}
        & Corticospinal tract (CST) \\
        \cline{2-2}
        & \makecell{Corticostriatal fibers (ST): \\ Fronto-orbital-striatal (ST\_FO), Prefrontal-striatal (ST\_PREF), \\ Premotor-striatal (ST\_PREM),  Precentral-striatal (ST\_PREC), \\ Postcentral-striatal (ST\_POSTC), Parieto-striatal (ST\_PAR), \\
        Occipito-striatal (ST\_OCC)} \\
        \cline{2-2}
        & \makecell{Thalamocortical fibers: \\ Anterior thalamic radiations (ATR), Thalamo-prefrontal (T\_PREF), \\ Thalamo-premotor (T\_PREM), Thalamo-precentral (T\_PREC), \\ Superior thalamic radiation (STR), Thalamo-postcentral (T\_POSTC), \\ Thalamo-parietal (T\_PAR), Optic radiation (OR), Thalamo-occipital (T\_OCC)} \\
        \hline
        
        \multirow{5}{*}{\textit{Association tracts}} & Uncinate fasciculus (UF) \\
        \cline{2-2}
        & Frontal Aslant Tract (FAT) \\
        \cline{2-2}
        & Inferior fronto-occipital fascicle (IFO) \\
        \cline{2-2}
        & Middle longitudinal fascicle (MLF) \\
        \cline{2-2}
        & Inferior longitudinal fascicle (ILF) \\        
        \hline

        \multirow{3}{*}{\textit{Cerebellar tracts}} & Inferior cerebellar peduncle (ICP) \\
        \cline{2-2}
        & Middle cerebellar peduncle (MCP) \\
        \cline{2-2}
        & Superior cerebellar peduncle (SCP) \\
        \hline

    \end{tabular}
    \caption{White matter tracts used in this study for an auxiliary loss term, and for evaluation of the registration performance across methods. For a detailed description of these tracts please see \cite{calixto2025detailed}.}
    \label{tab:summary_tracts}
\end{table}

For the dHCP fetal subjects, we used dMRI data that had been preprocessed with the SHARD pipeline \cite{christiaens2021scattered}. We did not perform any further preprocessing on the data, except for resampling the dMRI volumes from an isotropic voxel size of 2 mm to 1.2 mm. The diffusion tensor was then computed using a standard weighted linear least squares method \cite{koay2006unifying}. Since the dHCP data varies in matrix size, we automatically padded or cropped the resampled volumes to a spatial matrix size of $128^3$ before applying the trained model.

The registration methods developed in this work require segmentation of the fetal brain to exclude other fetal and maternal anatomy. Most of the BCH data included in this study had been preprocessed previously, where the brain segmentation was performed using a semi-automatic method in ITK-SNAP \cite{yushkevich2016itk}. Specifically, an initial ellipsoid brain mask was place on the fetal brain. Then, an experienced annotator manually refined this mask. In our current preprocessing pipeline, we use our deep learning methods for this purpose \cite{faghihpirayesh2024fetal, snoussi2024haitch}. For the dHCP subjects, we used the fetal brain masks that are provided with the data release.


\subsection{Evaluation metrics}\label{ss:metrics}

The quality of affine and deformable registration was evaluated using multiple metrics: the Dice coefficient for the warped white matter tract segmentations, image-wise cross-correlation for the registered FA maps, average cosine similarity for the registered diffusion tensors, and the physical plausibility of the estimated deformation fields, measured as the percentage of voxels with a negative Jacobian determinant (NJD\%) in the region where the displacement is non-zero. These metrics were computed both between the template and subjects, as well as between pairs of subjects, and were restricted to the brain region.

Notably, the upper bound of the Dice is less than one. This is because, while the Dice is averaged across non-empty labels, some tract segmentations, particularly those with bilateral structures, may be incomplete in individual scans, with either the left or right structure missing. In contrast, the tract atlas always presents complete structures, resulting in a maximum Dice below one.

Image-wise cross-correlation (CC) between two scalar images is computed as
\begin{equation}\label{eq:cc}
\text{CC}(I_1, I_2) = \frac{\sum_{x \in \Omega} (I_1(x) - \bar{I}_1)(I_2(x) - \bar{I}_2)}{\sqrt{\sum_{x \in \Omega} (I_1(x) - \bar{I}_1)^2 \sum_{x \in \Omega} (I_2(x) - \bar{I}_2)^2}},
\end{equation}
where $I_1(x)$ and $I_2(x)$ represent the intensity values at position $x$ in the 3D images, and $\bar{I}_1$ and $\bar{I}_2$ are the mean intensity values of $I_1$ and $I_2$ across all voxels in $\Omega$, the space of the brain region.

In addition, we utilize the Tenengrad sharpness measure \cite{pertuz2013analysis}, which quantifies image sharpness as the sum of squared gradient magnitudes, to evaluate the clarity of the mean FA map obtained using all methods. A higher sharpness value indicates better alignment, where anatomical structures overlap well, resulting in a clearer mean image. In contrast, lower sharpness suggests residual misalignment, as inconsistencies between subjects cause anatomical details to blur and average out, reducing clarity.

\subsection{Baseline methods}\label{ss:baseline_methods}
Several state-of-the-art algorithms for spatiotemporal analysis of diffusion tensor images were implemented in this study, including classical instance optimization-based methods FSL \cite{Smith2006tract} and DTI-TK \cite{zhang2006deformable}, and learning-based techniques SynthMorph \cite{hoffmann2024anatomy} and VoxelMorph \cite{balakrishnan2019voxelmorph}.

\textbf{\textit{FSL TBSS}} (version 6.0.7.11), typically used for adult brain imaging, was adapted for this study to accommodate fetal brain imaging by incorporating fetal brain templates. Including the default configuration, the following algorithms were implemented:

\begin{enumerate}

    \item The default configuration, which uses the FMRIB58\_FA mean FA image in standard space as the target image for spatial normalization, and uses its standard skeleton for the projection of individual FA images (denoted as the -T option; see Section~\ref{ss:background_tbss}).
    \item The pipeline that finds the most ``typical'' subject in the group as the target image for spatial normalization (the -n option).
    \item The pipeline that allows to specify an FA image as the group-wise target image for spatial normalization (the -t option). For this, we utilized GA-specific fetal FA template from an existing fetal DTI template \cite{Khan2019fetal}, which we will refer to as CRL template in this paper.   
    \item Another variant of the -t option pipeline, where we used the age-matched template from a more recent fetal DTI template that was constructed using superior fetal diffusion MRI scans from the developing Human Connectome Project (dHCP) data (\url{https://gin.g-node.org/kcl_cdb/fetal_brain_mri_atlas}).
    \item A custom pipeline developed by combining the capabilities of DTI-TK and TBSS 
        (\url{https://dti-tk.sourceforge.net/pmwiki/pmwiki.php?n=Documentation.TBSS}). Briefly, FSL's FLIRT and FNIRT were used to register each subject's FA image to the GA-specific FA template as used in (3). Subsequently, projection and skeletonization were performed in the template space using FSL TBSS. The key distinction between this custom pipeline and the aforementioned integrated TBSS pipelines (1) - (4) lies in the final space used for spatial analysis. Regardless of the target image chosen, the integrated pipelines automatically take both the target and normalized images into $1 \times 1 \times 1$ mm MNI152 space through registration to the FMRIB58\_FA template, which is constructed using adult images. In contrast, the custom pipeline can maintain the analysis within the GA-specific template space, avoiding potential issues associated with registering fetal brain images to an adult brain template. 

\end{enumerate}

\textbf{\textit{DTI-TK}} (version 2.3.1). The DTI-TK algorithm approximates the underlying deformable transformation between diffusion tensor images through piece-wise affine registration \cite{zhang2006deformable}. It has demonstrated superior performance over the FA-based registration employed in the original TBSS pipeline, owing to its ability to generate a group-specific DTI template by performing group-wise registration directly on tensor images \cite{Bach2014methodological, Yushkevich2009Structure, Keihaninejad2012importance}. Using this approach, we computed weekly GA-specific diffusion tensor templates. These templates were essentially correspondent with the FA templates from the CRL atlas mentioned above, and served as the group-wise target for all methods compared in this study.

\textbf{\textit{VoxelMorph}} represents a deep learning-based approach for deformable brain MR image registration \cite{balakrishnan2019voxelmorph}. It has demonstrated improved registration accuracy by incorporating the spatial correspondence of anatomical segmentations as an auxiliary loss term. To fully leverage this design, we included the Dice coefficient for available white matter tract segmentations, in addition to the original FA similarity measured using normalized cross-correlation with a window size of 9, and a regularization term. Since VoxelMorph is designed specifically for deformable registration and requires affine-aligned input images, we employed \textbf{\textit{SynthMorph}} \cite{hoffmann2024anatomy}—a state-of-the-art affine registration framework—as the affine component for VoxelMorph. 

\subsection{Implementation details}
We adopted the same hyperparameters that were optimized and reported in the literature for our implementation. The learning-based algorithms were trained on a workstation equipped with an NVIDIA RTX A6000 GPU (48 GB memory) and an AMD Ryzen Threadripper PRO 5965WX 24-core CPU. For FetDTIAlign, the loss function weight $\lambda$ was set to 10 for affine registration and 100 for deformable registration, with $\gamma$ set to 100. We used the Adam optimizer \cite{kingma2014adam} with an initial learning rate of $1 \times 10^{-4}$ and a batch size of 1.

\section{Results}

First, we present illustrative results from the GA23 group ($n=7$), focusing on optimizing the configuration of the classical approaches. Subsequently, using the optimal configuration, we summarize the quantitative evaluation results for all methods across the GA25, GA30, and GA35 groups. Across the figures in the presented study, differences in brain size and background area reflect natural brain development, not cropping or resizing.

\subsection{Illustrative results from the GA23 group}

\begin{figure*}[!h]
\caption{Spatial normalized FA images of the GA23 group using \textit{FSL TBSS-(1)} pipeline. The final results were automatically transformed to MNI152\_1mm space, and visualized in the axial, coronal, and sagittal planes.}
\centering
\includegraphics[width=0.9\textwidth]{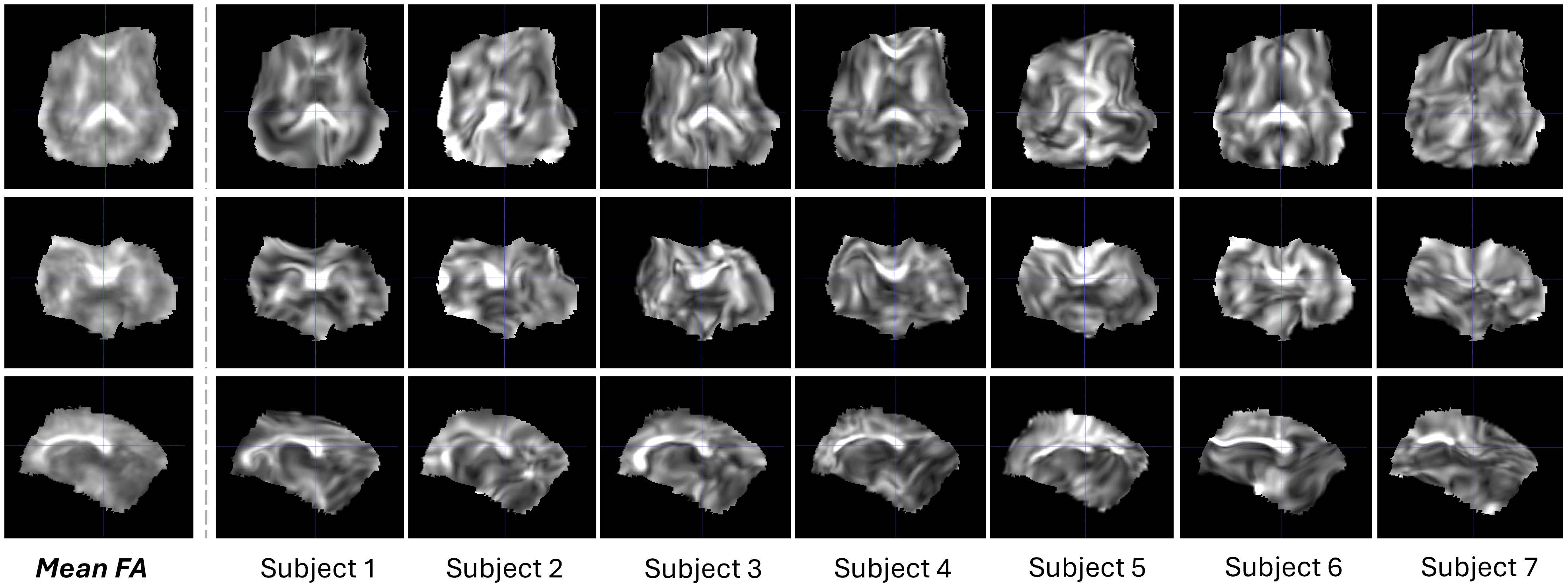}
\label{fig:registration_fsl_tbss_T}
\end{figure*}

\begin{figure*}[ht]
\caption{Spatial normalized FA images of the GA23 group using \textit{FSL TBSS-(4)} pipeline. The final results were automatically transformed to MNI152\_1mm space, and visualized in the axial, coronal, and sagittal planes.}
\centering
\includegraphics[width=0.9\textwidth]{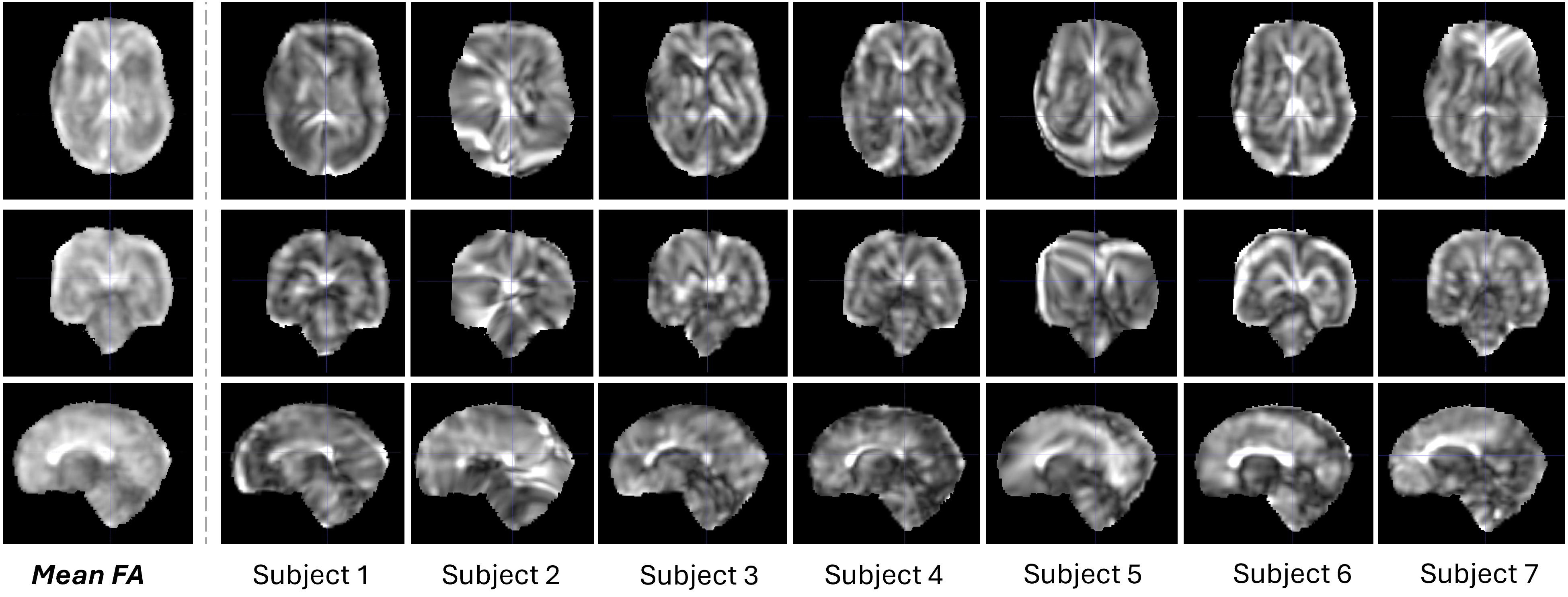}
\label{fig:registration_fsl_tbss_t_kcl}
\end{figure*}

\begin{figure*}[ht]
\caption{Spatial normalized FA images of the GA23 group using \textit{FSL TBSS-(5)} setting. The final results remained in the GA23-specific FA template space of isotropic 1.2mm resolution, and visualized in the axial, coronal, and sagittal planes.}
\centering
\includegraphics[width=0.9\textwidth]{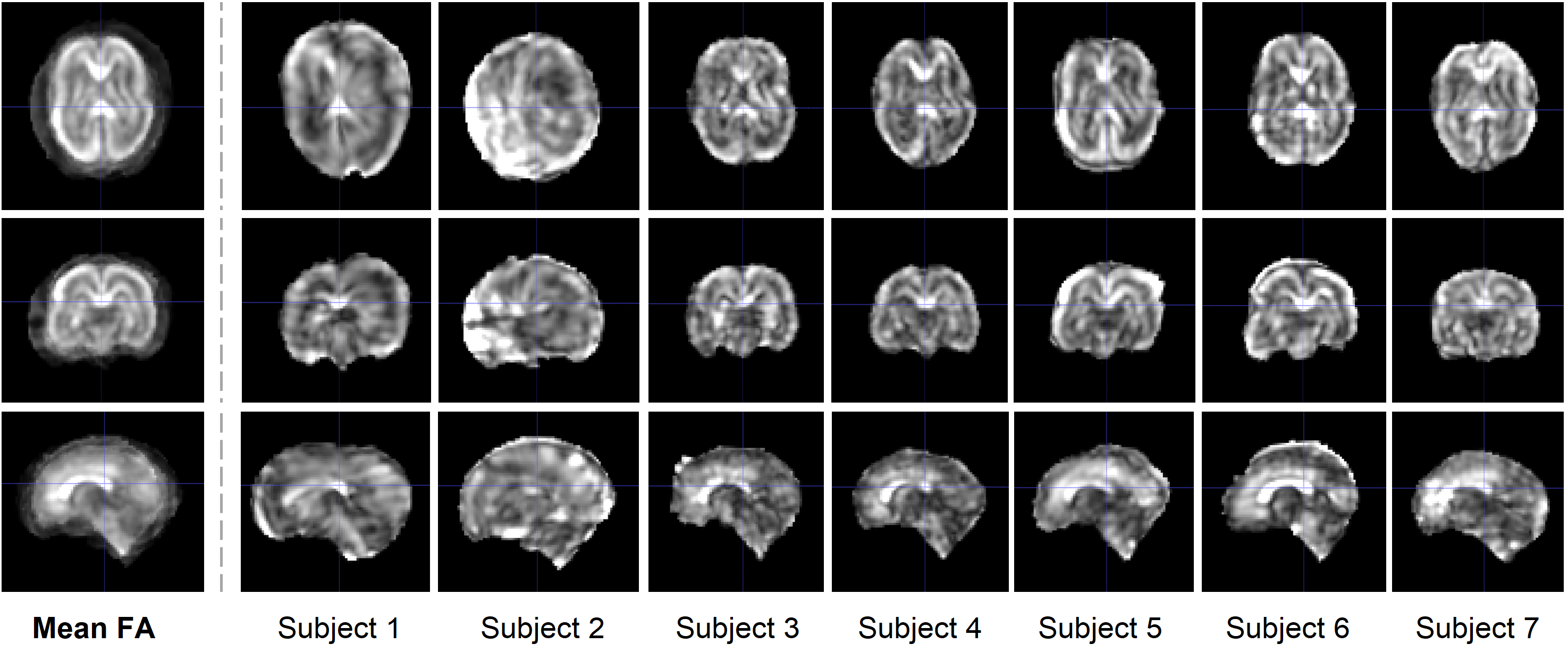}
\label{fig:registration_fsl_tbss_t_manual_inhome}
\end{figure*}

Qualitatively, using the default settings of FSL TBSS was not able to appropriately register the data from the GA23 group, regardless of the target image used for spatial normalization, as is shown in Figure \ref{fig:registration_fsl_tbss_T} for \textit{FSL TBSS-(1)} using the default adult template in MNI152 standard space, and Figure \ref{fig:registration_fsl_tbss_t_kcl} for \textit{FSL TBSS-(4)} using very high-quality KCL fetal template (0.5mm resolution). Similarly, \textit{FSL TBSS-(2)} and \textit{FSL TBSS-(3)} both failed when automatically aligning the group target image to standard MNI152\_1mm space for projection and skeletonization. Lastly, \textit{FSL TBSS-(5)}, which remained in the group-specific template space, produced reasonable results on good-quality FA images (Figure \ref{fig:registration_fsl_tbss_t_manual_inhome}: subjects 4), but failed to handle FA images with localized artifacts, such as hyper-intensity confined to the left hemisphere (Figure \ref{fig:registration_fsl_tbss_t_manual_inhome}: subject 2). Based on these observations, we selected configuration \textit{FSL TBSS-(5)} to represent the results of FSL TBSS in the following quantitative evaluations.  

\begin{figure*}[ht]
\caption{Results of the GA23 group using DTI-TK with the CRL fetal DTI template as the target image for spatial normalization, followed by FA computation from the spatially normalized tensor images. The final results remained in the GA23-specific DTI template space of isotropic 1.2mm resolution, and visualized in the axial, coronal, and sagittal planes.}
\centering
\includegraphics[width=0.9\textwidth]{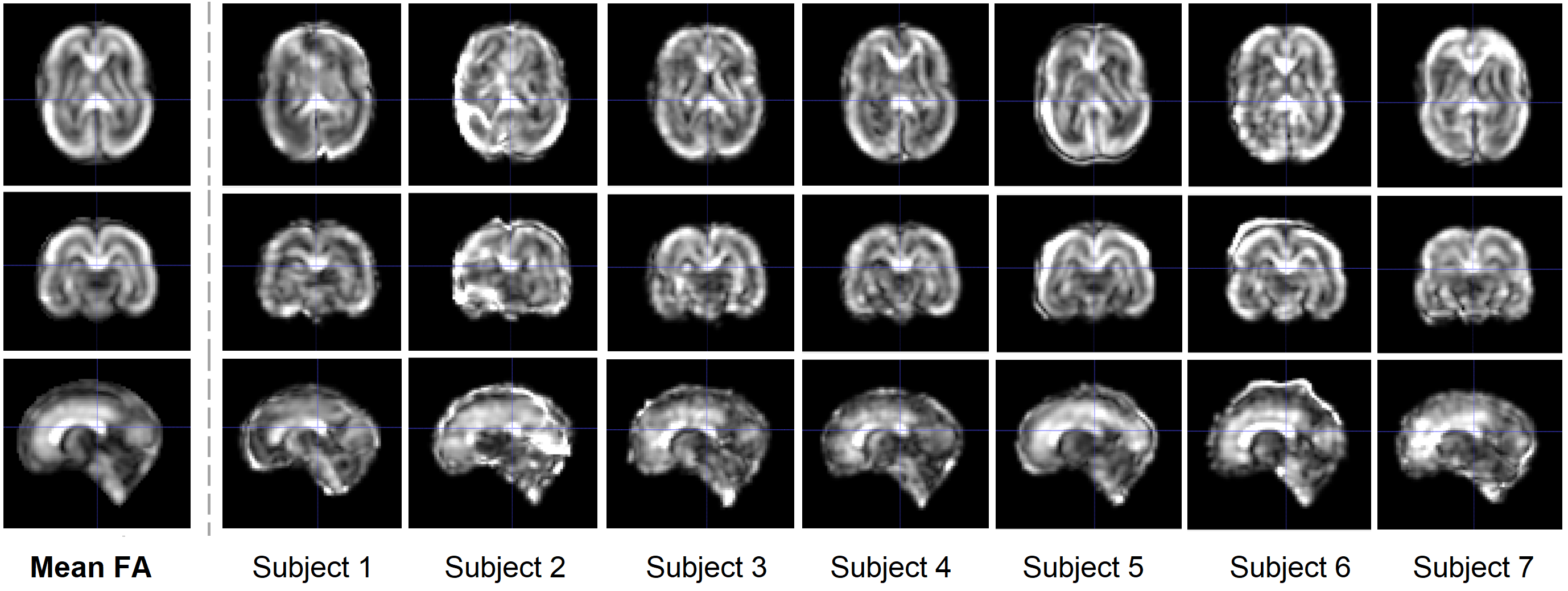}
\label{fig:registration_dtitk_tbss_inhome_1.2mm}
\end{figure*}

\begin{figure*}[ht]
\caption{Results of the GA23 group using FetDTIAlign with the CRL fetal DTI and FA template as the target image for spatial normalization. The final results remained in the GA23-specific template space of isotropic 1.2mm resolution, and visualized in the axial, coronal, and sagittal planes.}
\centering
\includegraphics[width=0.9\textwidth]{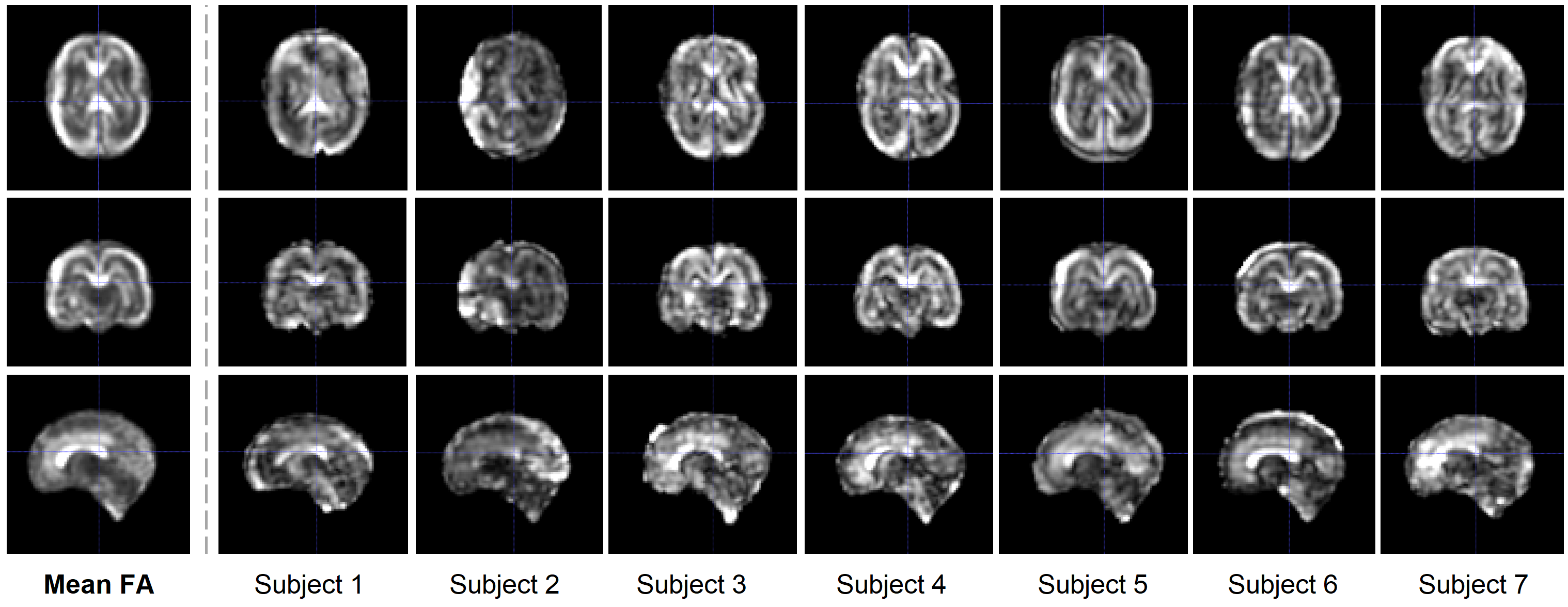}
\label{fig:registration_proposed_GA23_inhome_1.2mm}
\end{figure*}

In cases where direct registration using FA images failed, diffusion tensor-based registration followed by FA computing had less issue, resulting in better overall alignment with the template and preserving plausible brain shape. This improvement is best demonstrated in the results for Subject 2, comparing FSL TBSS (Figure~\ref{fig:registration_fsl_tbss_t_manual_inhome}) and \textit{DTI-TK} (Figure~\ref{fig:registration_dtitk_tbss_inhome_1.2mm}). The proposed algorithm produced similarly robust alignment for Subject 2, but visually better aligned the left frontal region and without stretching the spike for Subject 6 as seen in the sagittal plane (Figure~\ref{fig:registration_proposed_GA23_inhome_1.2mm}). 

\subsection{Registration quality}

\begin{figure}[h]
\caption{Original, rotated, and reoriented DTI templates for a 23-week gestational age (GA), using a synthesized affine transformation and the custom reorientation layer in FetDTIAlign.}
\centering
\includegraphics[width=0.5\textwidth]{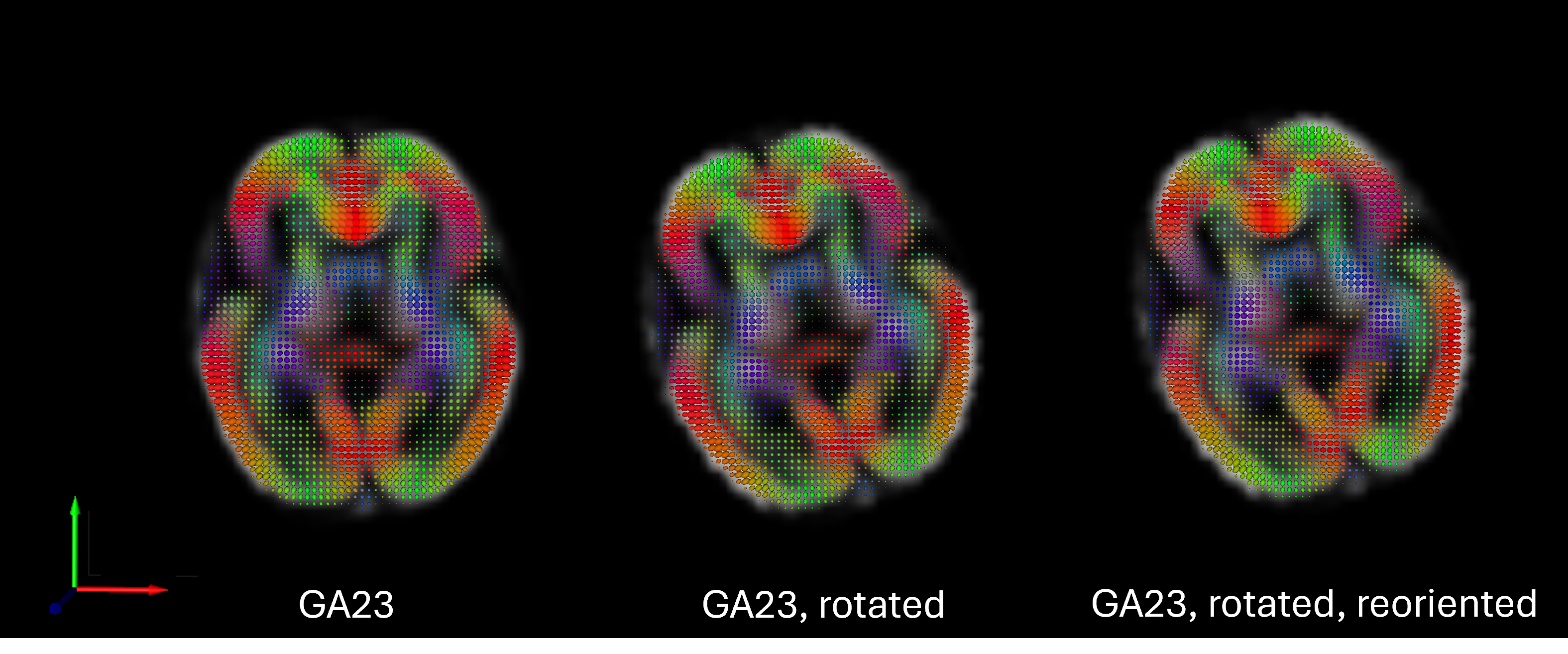}
\label{fig:check_reo}
\end{figure}

Using the synthesized affine transformations, we visually verified the reliability of both the data augmentation process and the custom reorientation layers in the proposed framework. An example is shown in Figure~\ref{fig:check_reo}.

Quantitatively, we focus on interpreting the results of the GA25, GA30, and GA35 groups, across the methods and thirty-four white matter tract labels (Table~\ref{tab:registration_table_interSubjs}).

\begin{sidewaystable}
    \centering
    \scriptsize
    
    \begin{tabular}{c c ||c|c|c||c|c|c||c|c|c}
        \multirow{2}{*}{} & \multirow{2}{*}{} &  \multicolumn{3}{c||}{GA25 ($n=7$; 36 pairs}) & \multicolumn{3}{c||}{GA30 ($n=11$; 78 pairs}) & \multicolumn{3}{c}{GA35 ($n=4$; 15 pairs}) \\ 
        
           &  & Dice \textuparrow & CC \textuparrow & NJD(\%) \textdownarrow & Dice \textuparrow & CC \textuparrow & NJD(\%) \textdownarrow & Dice \textuparrow & CC \textuparrow & NJD(\%) \textdownarrow  \\ 
        \hline
        \multicolumn{2}{c||}{Initial} & $0.48 \pm 0.083$ & $0.20 \pm 0.084$ & - & $0.49 \pm 0.058$ & $0.16 \pm 0.081$ & - &$0.53 \pm 0.040$ & $0.20 \pm 0.077$ & - \\ 
        \hline
        \hline
        \multirow{4}{*}{\textbf{Affine}} &   \textit{FSL} &   $0.60\pm0.034$ &   $0.33\pm0.100$ &   - &   $0.56 \pm 0.039$ &   $0.24 \pm 0.073$ &   - & $0.58 \pm 0.025$ &  $0.26  \pm 0.057$ &   - \\     
        \cline{2-11}
        
        & \textit{DTI-TK} &   $0.38 \pm 0.130$ &   $0.39 \pm 0.089$ &   - &   $0.45 \pm 0.080$ &   $0.26 \pm 0.100$ &   - &  $0.47 \pm 0.070$ &  $0.23 \pm 0.064$ &   - \\ 
        \cline{2-11}
        
        & \textit{SynthMorph} &   $0.64 \pm 0.028$ &   $0.50 \pm 0.089$ &   - &   $0.60 \pm 0.031$ &   $0.35 \pm 0.088$ &   - &   $0.58 \pm 0.023$ &   $0.27 \pm 0.064$ &   - \\
        \cline{2-11}
        
        & \textit{FetDTIAlign} &  $\mathbf{0.67 \pm 0.022}^{**}$ &  $\mathbf{0.52 \pm 0.086}^{**}$ &   - & $\mathbf{0.62 \pm 0.032}^{**}$ & $\mathbf{0.36 \pm 0.091}^{**}$ &   - & $\mathbf{0.62 \pm 0.010 }^{*}$ & $\mathbf{0.29 \pm 0.062}^{*}$ &   - \\ 
        \hline
        \hline        
        
         \multirow{4}{*}{\textbf{Non-linear}} & \textit{FSL} & $0.50\pm0.092$ & $0.24\pm0.110$ & $3.2 \pm 1.2$ & $0.50 \pm 0.075$ & $0.23 \pm 0.120$ & $6.4 \pm 2.5$ & $0.56 \pm 0.033$ & $0.26 \pm 0.057$ & $9.6 \pm 1.3$ \\ 
        \cline{2-11}

         & \textit{DTI-TK} & $0.39 \pm 0.130$ & $0.69\pm0.059$ & $\mathbf{0.00 \pm 0.00}^{*}$ & $0.46 \pm 0.095$ & $0.57 \pm 0.094$ & $\mathbf{0.00 \pm 0.00}^{*}$ & $0.50 \pm 0.080$ & $0.36 \pm 0.072$ & $\mathbf{0.00 \pm 0.00}^{*}$ \\          
        \cline{2-11}
        
        & \textit{VoxelMorph} & $0.68 \pm 0.022$ & $\mathbf{0.91 \pm 0.024}^{**}$ & $0.027 \pm 0.017$ & $0.64 \pm 0.031$ & $\mathbf{0.86 \pm 0.052}^{**}$ & $0.10 \pm 0.052$ & $0.63 \pm 0.011$ & $\mathbf{0.73 \pm 0.054}^{**}$  & $0.096 \pm 0.022$ \\                
        \cline{2-11}
        
        & \textit{FetDTIAlign} & $\mathbf{0.75 \pm 0.019}^{**}$ & $0.80 \pm 0.050$ & $ 6.5 \pm 5.4 (\times 10^{-5})$ & $\mathbf{0.70 \pm 0.032}^{**}$ & $0.67 \pm 0.085$ & $2.8 \pm 2.5 (\times 10^{-3})$ & $\mathbf{0.68 \pm 0.008}^{**} $ & $0.65 \pm 0.053$ & $0.038 \pm 0.013$ \\          
        \hline        
    \end{tabular}
    \caption{Registration quality across the methods and age groups, as indicated by metrics of the average Dice coefficient between the warped masks of 60 white matter tracts (Dice), cross-correlation between registered FA maps (CC; Eq.~\ref{eq:cc}), and the smoothness of the estimated deformation fields (NJD). Metrics were computed in an inter-subjects manner, and only within a valid region, i.e., the Dice is averaged over available tracts in individual scans; CC is averaged over only the voxels within the target brain; NJD(\%) is a percentage over voxels where any of the three displacement components are greater than 0. \textbf{Bold} value indicate a higher or lower value compared with the others. ** denotes a p-value $\ll 0.01$, and * denotes a p-value $<0.05$ for the paired t-test}.
    \label{tab:registration_table_interSubjs}
     
\end{sidewaystable}

\begin{figure*}[!htbp]
\caption{Visualization of the spatial correspondence of left and right Striato-Parietal tract (ST\_PAR) for the seven subjects in GA25 group that were warped and overlaid (in gray) with the tract on the target atlas (in red) using the affine transformation estimated by FSL, DTI-TK, and FetDTIAlign. Arrows point to the regions where different methods show markedly different registration accuracy at the tract boundary.}
\centering
\includegraphics[width=0.9\textwidth]{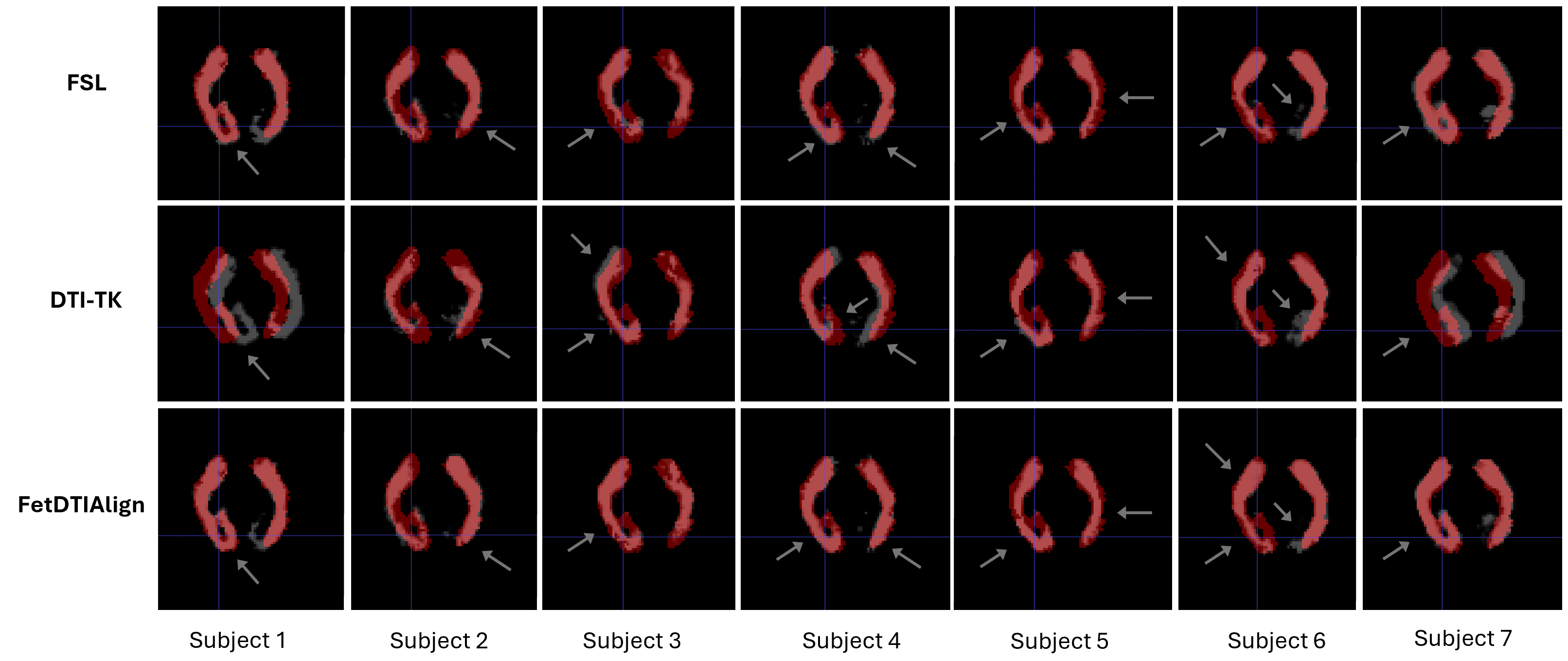}
\label{fig:st_par_GA25}
\end{figure*}

\begin{figure*}[ht]
\caption{Visualization of the spatial correspondence of left and right Inferior Longitudinal Fascicle (ILF) for the seven subjects in GA25 group that were warped and overlaid (in gray) with the tract on the target atlas (in red) using the affine transformation estimated by FSL, DTI-TK, and FetDTIAlign. Arrows point to regions where different registration methods have resulted in markedly different alignment accuracy at the tract boundary.}
\centering
\includegraphics[width=0.9\textwidth]{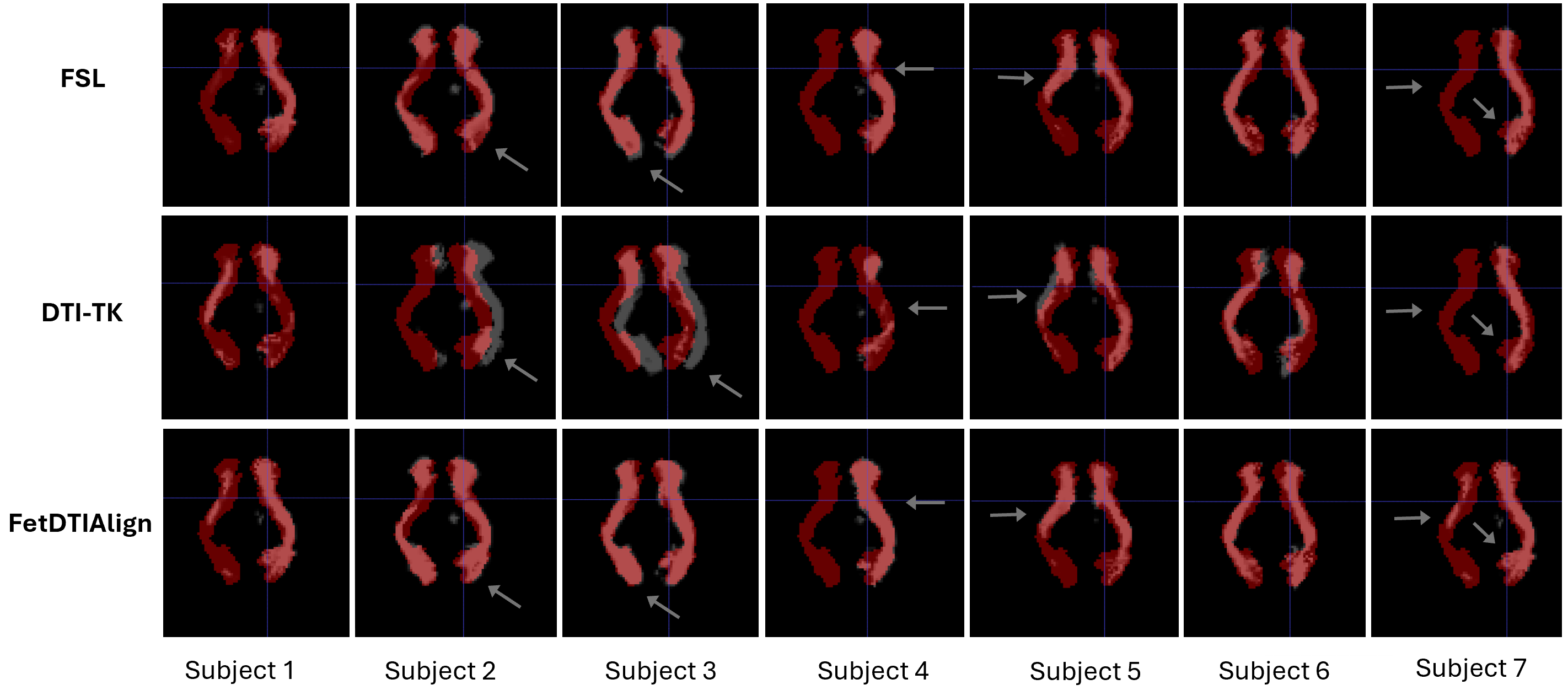}
\label{fig:ilf_GA25}
\end{figure*}

For affine registration, the proposed approach achieved the highest average Dice coefficient for warped white matter tracts across all three age groups, followed by SynthMorph and FSL. DTI-TK, however, produced lower Dice coefficients than the initial values prior to affine registration. Figure~\ref{fig:st_par_GA25} visualizes the spatial correspondence of the warped left and right Striato-Parietal tracts (ST\_PAR) overlaid (in gray) with the tract atlas (in red) using transformations estimated by FSL, DTI-TK, and the proposed method for the GA25 group. Together with the visualization of the Inferior Longitudinal Fascicle (ILF) (Figure~\ref{fig:ilf_GA25}), these views highlight the challenges of the task, such as the individual variances and the incomplete segmentation in individual scans, even with accurate alignment (e.g., Subject 3; Proposed approach). For the GA25 group, the lower Dice coefficient for DTI-TK was primarily due to translation errors observed in Subject 1 and Subject 7, as well as slight rotational errors in Subject 3 and Subject 4. The results of FSL generally aligned the ST\_PAR tract accurately, with minor misalignments along the boundaries. These residual misalignments were further corrected by the proposed method. The results for the proposed method are visually better than those for the competing methods, with the residual non-overlapped regions largely reflecting intrinsic inconsistencies between individual anatomy and the atlas. Attempting to further align these regions for this tract specifically would likely introduce errors in other tracts and lead to an implausible brain shape, rather than a genuine improvement.

\begin{figure*}[!htbp]
\caption{Initial FA images and the results of affine registration for the GA25 group using FSL, DTI-TK, and FetDTIAlign. The first column displays the target template for the initial images, as well as the Mean FA images averaged over the registered FA images for each respective method. The results are presented with isotropic 1.2 mm spacing and the same matrix size across all age groups to facilitate comparison. Arrows point to some of the regions where differences in registration accuracy between different methods have resulted in varying mean FA quality and sharpness. The Tenengrad sharpness of the Mean FA map obtained by the three methods is $5.98 \times 10^{4}$ (FSL), $7.30\times 10^{4}$ (DTI-TK), and $8.47\times 10^4$ (FetDTIAlign).}
\centering
\includegraphics[width=0.9\textwidth]{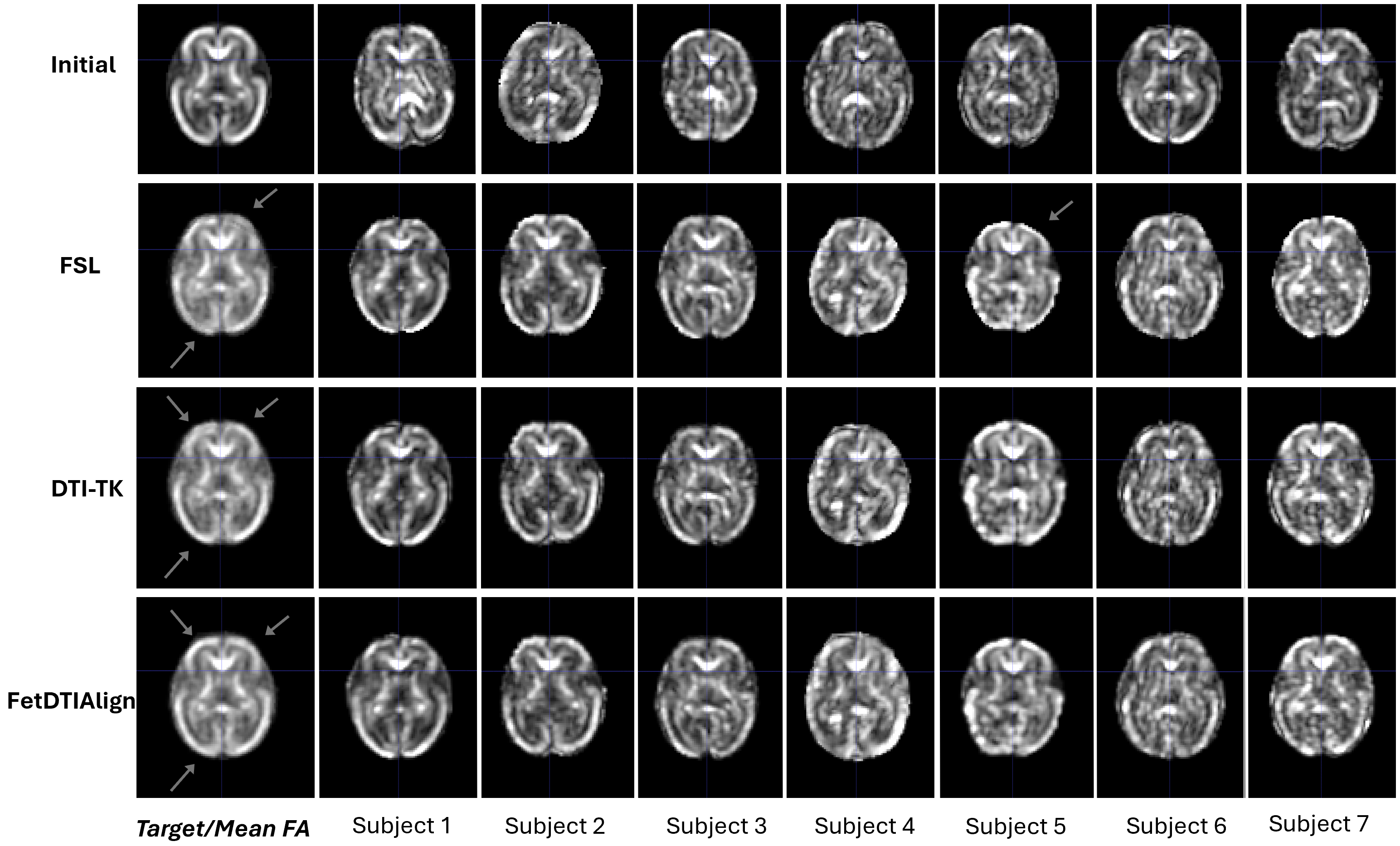}
\label{fig:fa_affine_GA25}
\end{figure*}

\begin{figure*}[!htbp]
\caption{Initial FA images and the results of affine registration for the GA35 group using FSL, DTI-TK, and FetDTIAlign. The first column displays the Mean FA images averaged over the initial or registered FA images for each respective method. The results are presented with isotropic 1.2 mm spacing and the same matrix size across all age groups to facilitate comparison. Arrows point to regions with misalignment, characterized by blurring and outward-shifted edges at the brain boundary for the Mean FA map. The Tenengrad sharpness of the Mean FA map obtained by the three methods is $1.65 \times 10^{5}$ (FSL), $2.64\times 10^{5}$ (DTI-TK), and $2.77\times 10^5$ (FetDTIAlign).}
\centering
\includegraphics[width=0.9\textwidth]{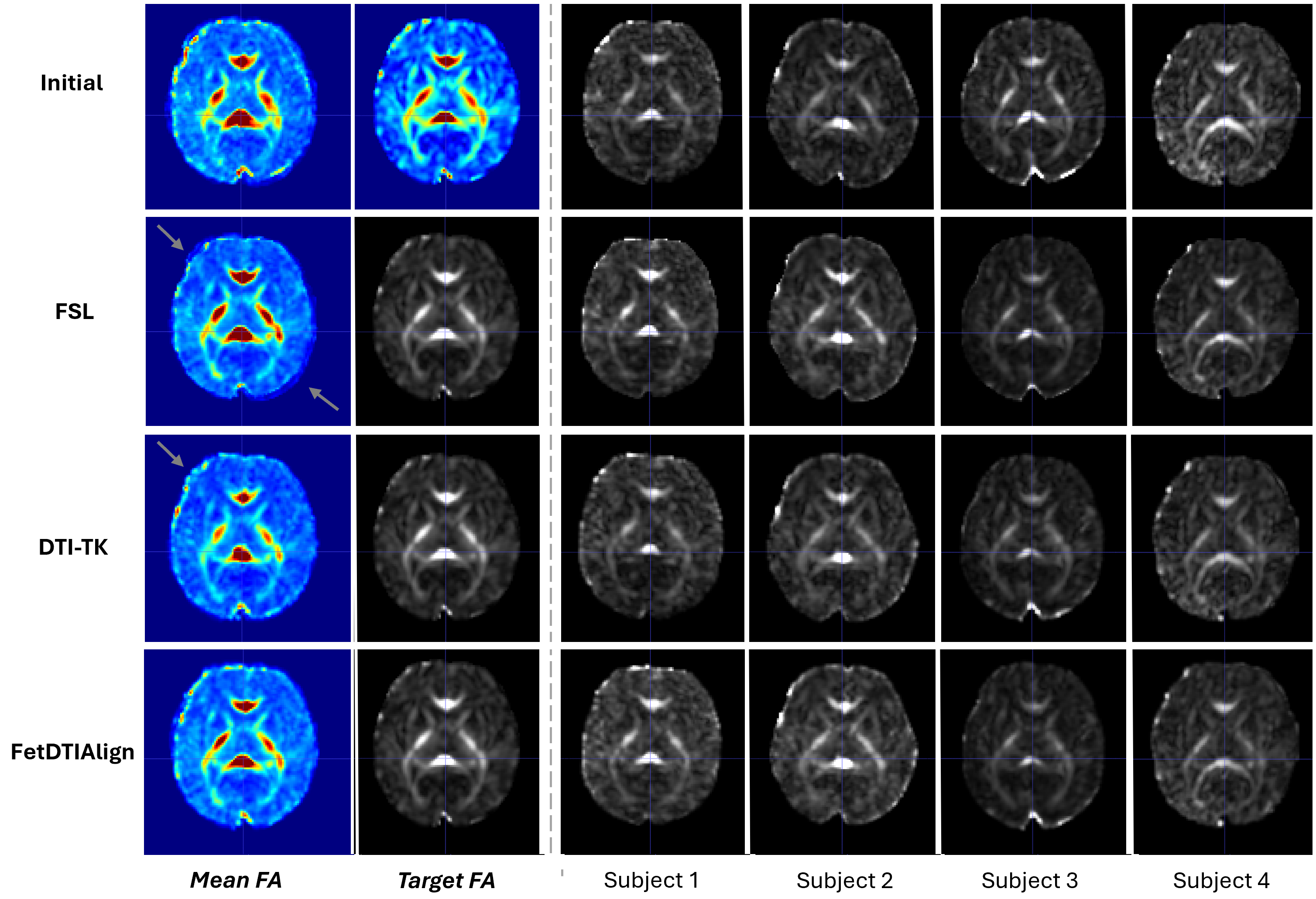}
\label{fig:fa_affine_GA35}
\end{figure*}

Visually, the center of mass of the initial FA images and the target template were aligned after diffusion MRI preprocessing and manual selection for the GA25 group. As a result, the affine registration for this group primarily involved non-translational transformations (Figure~\ref{fig:fa_affine_GA25}). The sharpness of the Mean FA image, computed from spatially normalized FA images, provides a visual reflection of alignment, complementing the quantitative measures. Higher sharpness indicates better alignment, where anatomical structures overlap well, leading to a clear mean image. In contrast, blurriness suggests residual misalignment, as inconsistencies between subjects cause anatomical details to average out, reducing clarity. The Mean FA image produced by the proposed method was the sharpest and most closely resembled the target template, indicating the best overall alignment. Notably, the target template is derived from the deformable results of DTI-TK, which itself was built on the affine registration results shown in the third row. FSL tended to produce an over-compression (Subject 5), or left-right compression of the frontal lobe compared with the target template (Mean FA), even when the initial moving image was closer to the target shape than the registered result (Subject 7). This observation is in line with the Tenengrad sharpness of the Mean FA map for the GA25 group obtained by the three methods, which is $5.98 \times 10^{4}$ (FSL), $7.30\times 10^{4}$ (DTI-TK), and $8.47\times 10^4$ (FetDTIAlign).

The average cross-correlation between affine registered FA images, computed using Eq.~\ref{eq:cc}, was highest for the proposed approach across all three brain development stages, followed by SynthMorph and DTI-TK for the GA25 and GA30 groups, while for the GA35 group FSL achieved a higher cross-correlation than DTI-TK (Table~\ref{tab:registration_table_interSubjs}). Overall, when comparing performance metrics across the three age groups, all methods showed higher FA similarity in the younger age groups, and additionally, showed higher tract correspondence for FSL, SynthMorph, and the proposed approach. Learning-based methods consistently outperformed classical methods, with the smallest performance difference observed in the GA35 group, where SynthMorph obtained similar Dice coefficients and FA cross-correlation to FSL. The initial images in the GA35 group were better aligned compared to those of younger age groups, likely due to the larger brain size and reduced free space, limiting variability in initial positioning (Figure~\ref{fig:fa_affine_GA35}). However, as brain development progresses, the increasingly complex geometries and greater individual variability in older brains pose additional challenges for image registration. The Tenengrad sharpness of the Mean FA map for the GA35 group (Figure~\ref{fig:fa_affine_GA35}) obtained by the three methods is $1.65 \times 10^{5}$ (FSL), $2.64\times 10^{5}$ (DTI-TK), and $2.77\times 10^5$ (FetDTIAlign). Across all methods, the mean FA maps for the GA35 group are sharper than those for the GA25 group, reflecting the ongoing myelination during development.

\begin{figure*}[!ht]
\caption{Initial FA images and the results of deformable registration for the GA25 group using FSL, DTI-TK, VoxelMorph (VXM), and FetDTIAlign. The first column displays the target template for the initial images, as well as the Mean FA images averaged over the registered FA images for each respective method. The results are presented with isotropic 1.2 mm spacing and the same matrix size across all age groups to facilitate comparison. The Tenengrad sharpness of the Mean FA maps obtained with different methods is $4.86 \times 10^{4}$ (FSL), $8.35\times 10^{4}$ (DTI-TK), $1.02 \times 10^5$ (VXM), and $1.28\times 10^5$ (FetDTIAlign).}
\centering
\includegraphics[width=0.9\textwidth]{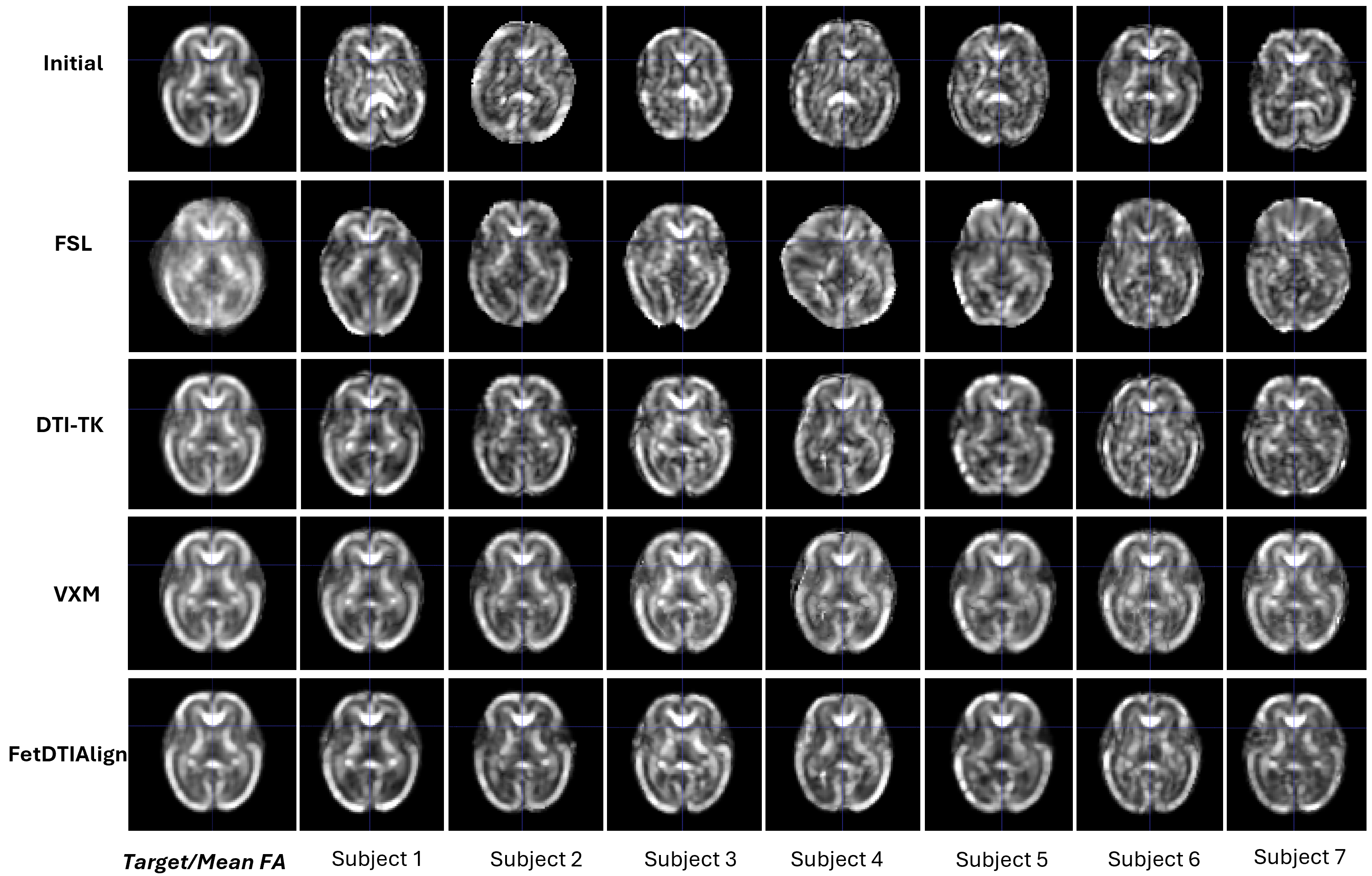}
\label{fig:fa_nonlinear_GA25}
\end{figure*}

For deformable registration, all methods used the affine-aligned results from their respective approaches as input, and the final results were resampled only once by composing the transformations from all stages (e.g., rigid, affine, and deformable stages for FSL). The proposed approach achieved the highest average Dice coefficient for warped white matter tracts across all three age groups, followed by VoxelMorph and FSL (Table~\ref{tab:registration_table_interSubjs}). With the addition of deformable registration, all methods improved their Dice coefficients as well as the cross-correlation, except for FSL, which were overall lower than its affine-aligned results across all three age groups. Deformable registration improved dramatically the average cross-correlation between registered FA images, in particular for VoxelMorph and for the younger age groups, increasing from 0.50 to 0.91 for the GA25 group and from 0.35 to 0.86 for the GA30 group. The improvements could be partly due to the fact that the use of normalized cross-correlation as the objective function optimized the method towards estimating transformations with a higher image-wise cross-correlation, especially when the kernel size is large. Also, the observed high cross-correlation accompanied with a higher percentage of negative Jacobian determinants and therefore a compromise of the physical plausibility of the deformation. In contrast, whereas the cross-correlations were low, DTI-TK achieved an ideally smooth deformation field (NJD=0.00\%) because it formulates the nonlinear deformation estimation as piece-wise affine estimations. The proposed method, on the other hand, produced a balance by achieving both high image similarity and smooth deformation. For instance, for the GA25 group, it obtained a cross-correlation of 0.80 compared with 0.69 for DTI-TK, and a lower percentage of negative Jacobian determinants (NJD) at 0.000065\%, compared with 0.027\% for VoxelMorph. 

Similar with the observation from the affine registration results, most deformable methods showed higher FA similarity in the younger age groups. DTI-TK showed increased tract correspondence in older age groups for both the affine and deformable registration, while with only marginal increase of the Dice coefficient with the addition of deformable registration. The Tenengrad sharpness of the mean FA map obtained using deformable registration methods (Figure~\ref{fig:fa_nonlinear_GA25}) is $4.86 \times 10^{4}$ (FSL), $8.35\times 10^{4}$ (DTI-TK), $1.02 \times 10^5$ (VXM), and $1.28\times 10^5$ (FetDTIAlign). The mean FA maps became sharper after deformable registration for all methods except FSL, which had a sharpness of $5.98 \times 10^{4}$ when aligned using affine registration.

These quantitative evaluations align with the visualizations (Figure~\ref{fig:fa_nonlinear_GA25}), where FSL, with lower cross-correlation of FA images, produced failed registrations in half of the GA25 group (Subjects 4–7). The observed effect is likely a result of the over-compression observed during the affine registration stage, which forced the deformable transformation, with its greater freedom, to over-stretch the frontal regions in order to compensate for the affine errors. In contrast, the results from VoxelMorph appeared to closely resemble the target template, as supported by its highest cross-correlation of FA, but perhaps too much so. Visual inspection of the registered images revealed several issues. First, all warped images appeared nearly identical to the target image, suggesting that the individual anatomical variations in the moving images were lost. Second, the deformation introduced `fake' structures - bright rim along the left and right boundaries of the warped brains. These structures did not present in some of the initial moving images (e.g., Subject 5, right frontal-temporal region), indicating that the registration process produced implausible deformations. Lastly, there appeared `white cracks' visible on the warped images (Subject 6 and 7), which were absent in both the moving and target images. These artifacts may be linked to unrealistic gradients in the deformation field, where the registration method attempts to achieve high similarity metrics at the expense of physical plausibility. Such issues may result from folding or tearing artifacts caused by large deformations and discontinuities.

\subsection{FA skeletons and all FA skeletonized}

\begin{figure*}[!ht]
\caption{Mean FA skeleton (green) and individual FA map projected onto the skeleton (red) using the deformable registration results from FSL, DTI-TK, VoxelMorph (VXM), and FetDTIAlign. The FA skeletons were computed using FSL TBSS (threshold of 0.2) based on the registered images, and serve as a reference for evaluating the accuracy of the registration methods. The results are presented in the sagittal plane with isotropic 1.2 mm spacing. Arrows point to regions with poor registration, leading to implausible brain shapes.}
\centering
\includegraphics[width=\textwidth]{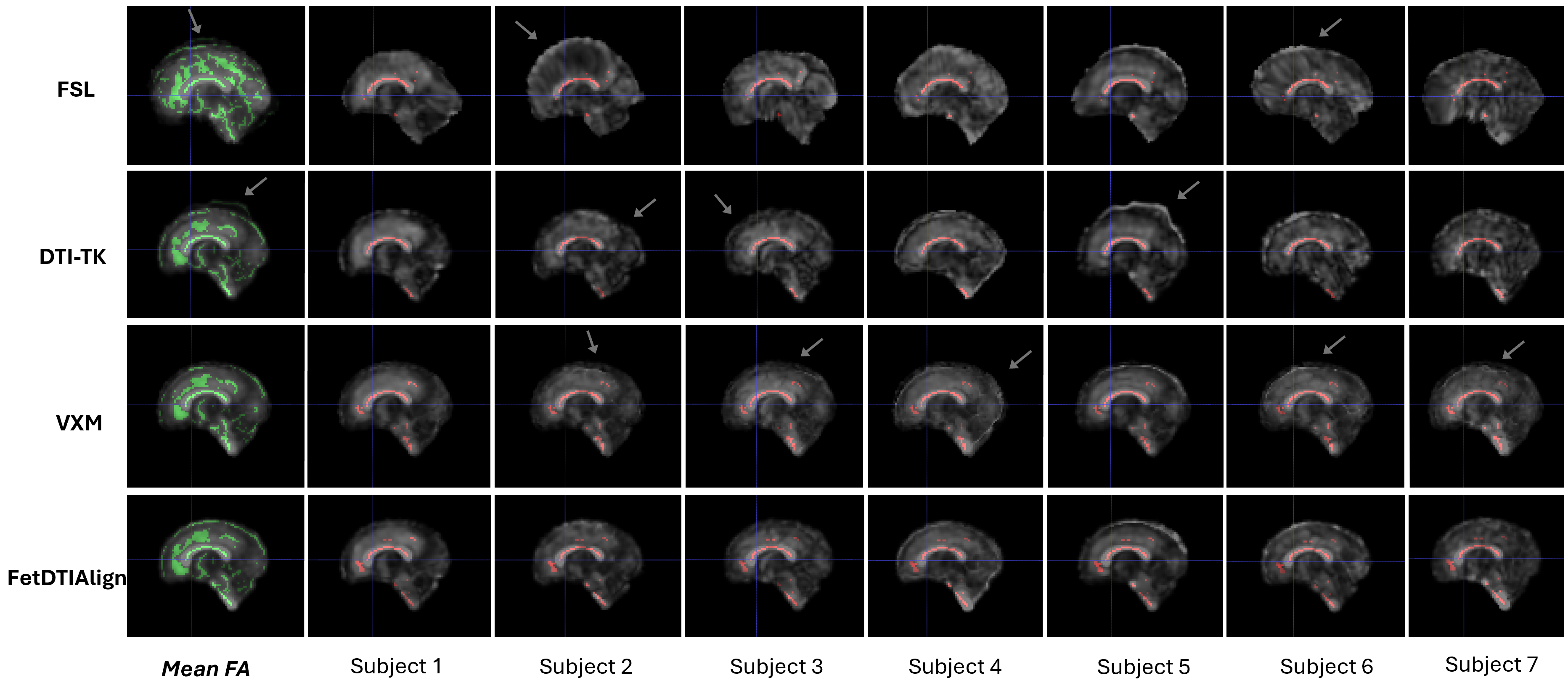}
\label{fig:skeleton_fa_nonlinear_GA25}
\end{figure*}

\begin{figure*}[h]
\caption{Mean FA skeleton (green) and individual FA map projected onto the skeleton (red) using the deformable registration results from FSL, DTI-TK, VoxelMorph (VXM), and FetDTIAlign. The FA skeletons were computed using FSL TBSS (threshold of 0.2) based on the registered images, and serve as a reference for evaluating the accuracy of the registration methods. The results are presented in the axial plane with isotropic 1.2 mm spacing.}
\centering
\includegraphics[width=0.9\textwidth]{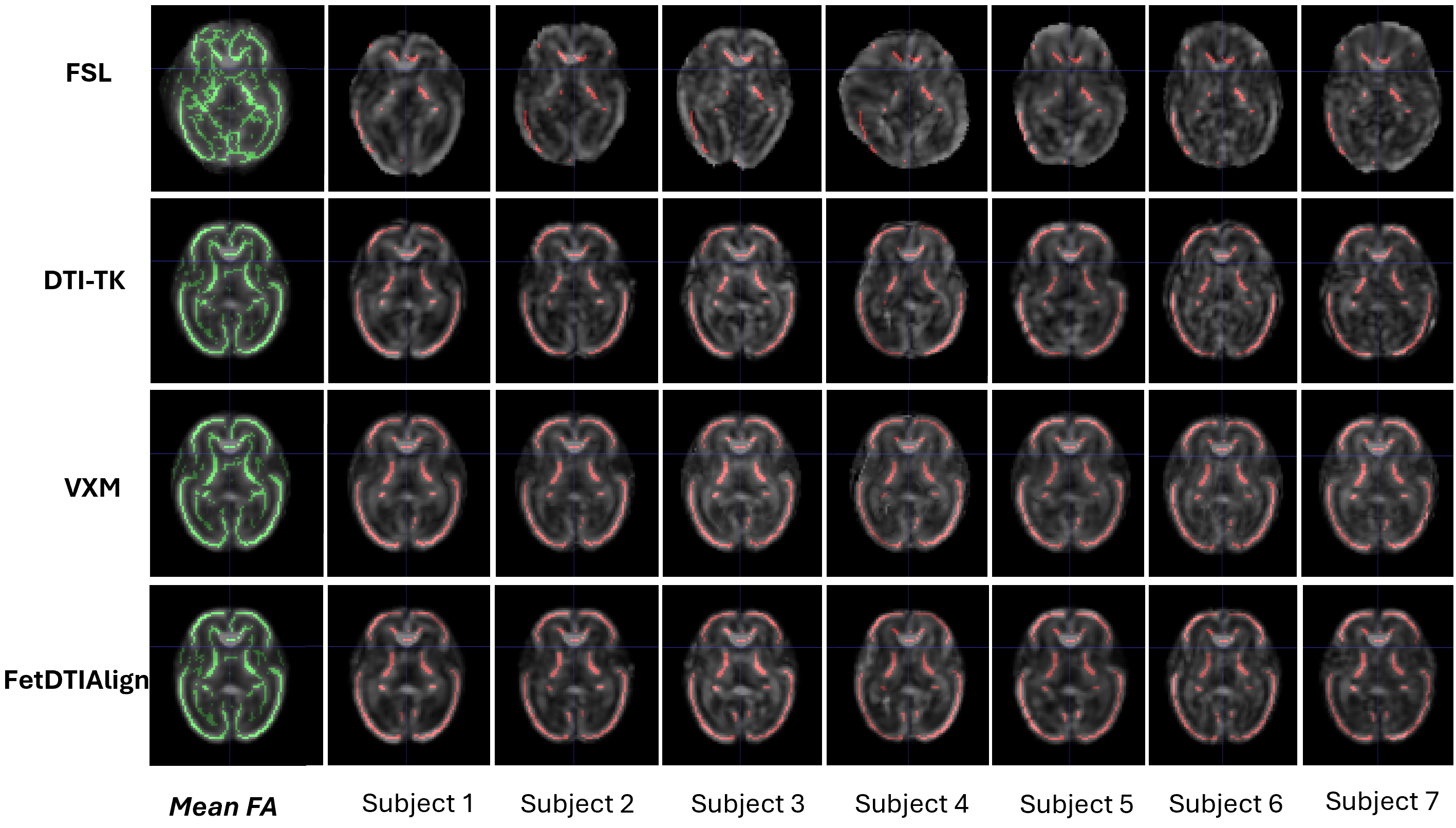}
\label{fig:skeleton_fa_nonlinear_GA25_axial}
\end{figure*}

\begin{figure*}[h]
\caption{Mean FA skeleton (green) and individual FA map projected onto the skeleton (red) using the deformable registration results from FSL, DTI-TK, VoxelMorph (VXM), and FetDTIAlign. The FA skeletons were computed using FSL TBSS (threshold of 0.2) based on the registered images, and serve as a reference for evaluating the accuracy of the registration methods. The results are presented in the coronal plane with isotropic 1.2 mm spacing.}
\centering
\includegraphics[width=0.9\textwidth]{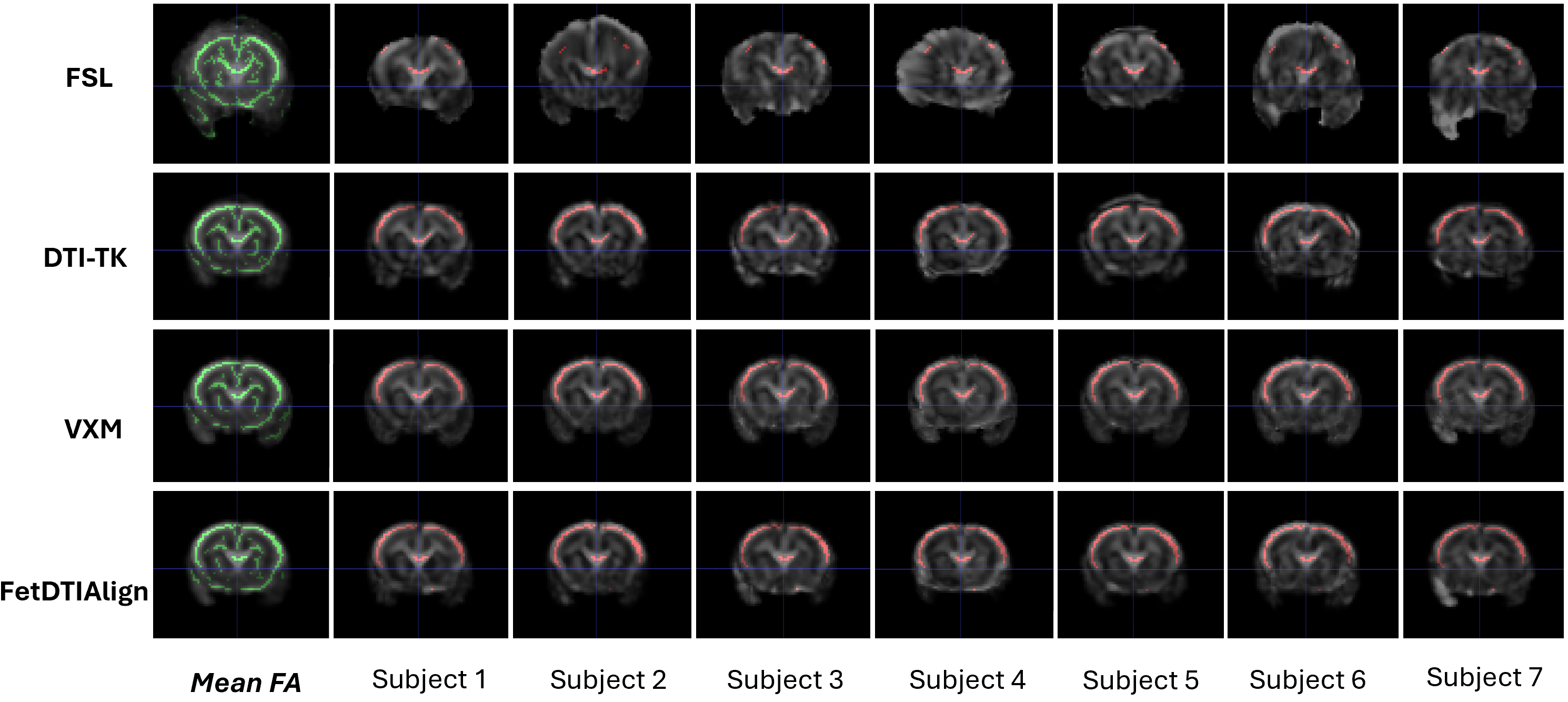}
\label{fig:skeleton_fa_nonlinear_GA25_coronal}
\end{figure*}

The FA skeletons were computed using FSL TBSS \cite{Smith2006tract} (with a threshold of 0.2; Section~\ref{ss:background_tbss}) from the spatially normalized FA images produced by all deformable registration methods, serving as a visual reference for evaluating inter-subject correspondence. Figure~\ref{fig:skeleton_fa_nonlinear_GA25} presents the Mean FA skeletons (in green) for the GA25 group, along with all FA images projected (in red) onto the skeleton using the spatially normalized images from the respective deformable registration methods. These images, shown in the sagittal plane, complement the deformable registration results displayed in the axial plane in Figure~\ref{fig:fa_nonlinear_GA25}. In general, the resulting FA skeletons showed a large agreement across the different registration approaches, particularly for DTI-TK, VoxelMorph, and the proposed method. FSL, however, failed to register most of the images accurately. Despite this, after thresholding and projection, only the skeletons (red) along the corpus callosum remained, making the results more comparable to the projections from the well-aligned methods. This suggests that the skeletonization procedure enhances robustness to misalignment. Results of the skeletons in the axial and coronal views are provided in Figures ~\ref{fig:skeleton_fa_nonlinear_GA25_axial} and ~\ref{fig:skeleton_fa_nonlinear_GA25_coronal}).

\begin{figure}[!ht]
\caption{Mean FA skeleton (green) across early brain development stages using the deformable registration results from the proposed methods. The results are presented with isotropic 1.2 mm spacing.}
\centering
\includegraphics[width=0.5\textwidth]{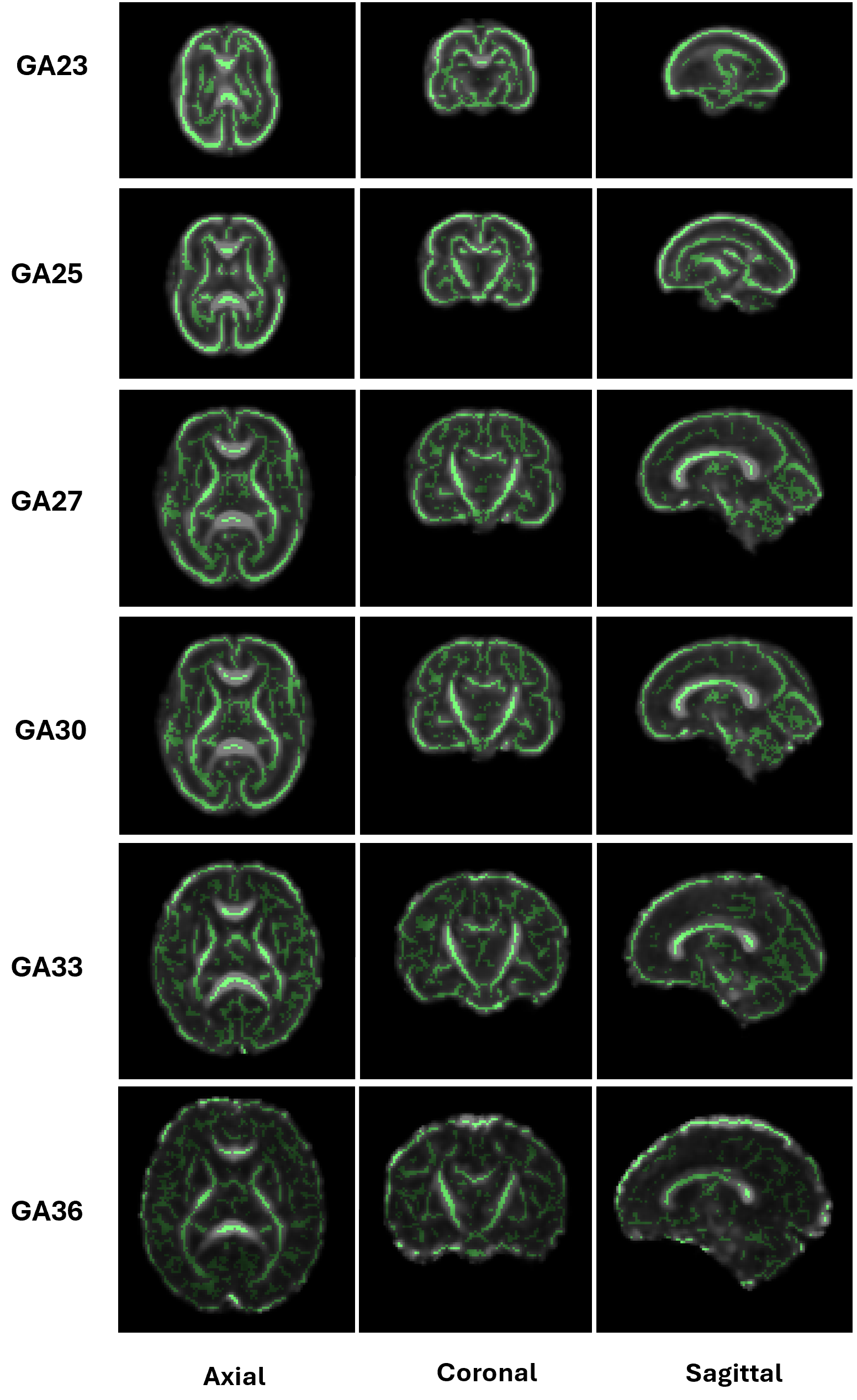}
\label{fig:skeleton_fa_nonlinear_acrossAge}
\end{figure}

In addition, we present the mean FA skeletons generated using the proposed approach across early brain development stages, providing insights into the brain growth trajectory (Figure~\ref{fig:skeleton_fa_nonlinear_acrossAge}).

\subsection{Generalizability on the external dHCP testing dataset}

To quantitatively evaluate the generalizability of the proposed method, we assessed its performance in registering dHCP fetal dMRI data to the closest GA-specific CRL fetal brain template. Paired t-tests were conducted to determine the statistical significance of the differences between the registration metrics.

We establish a ``reference'' similarity score for the dHCP dataset by applying a controlled translation to each subject image from this dataset. Specifically, we translated each subject image by one voxel (i.e., 1.2 mm) along the x, y, and z axes. We then computed the cross-correlation metric between the translated image and the original image. The resulting difference between the moved image and original image is due only to the small translations, thereby eliminating the variations due to the differences in acquisition protocols, preprocessing pipelines, or population characteristics. Hence, this reference similarity score serves as a baseline, reflecting the expected metric value for minor spatial misalignments without confounding factors.

\begin{figure}[!ht]
    \centering
    \includegraphics[width=0.5\linewidth]{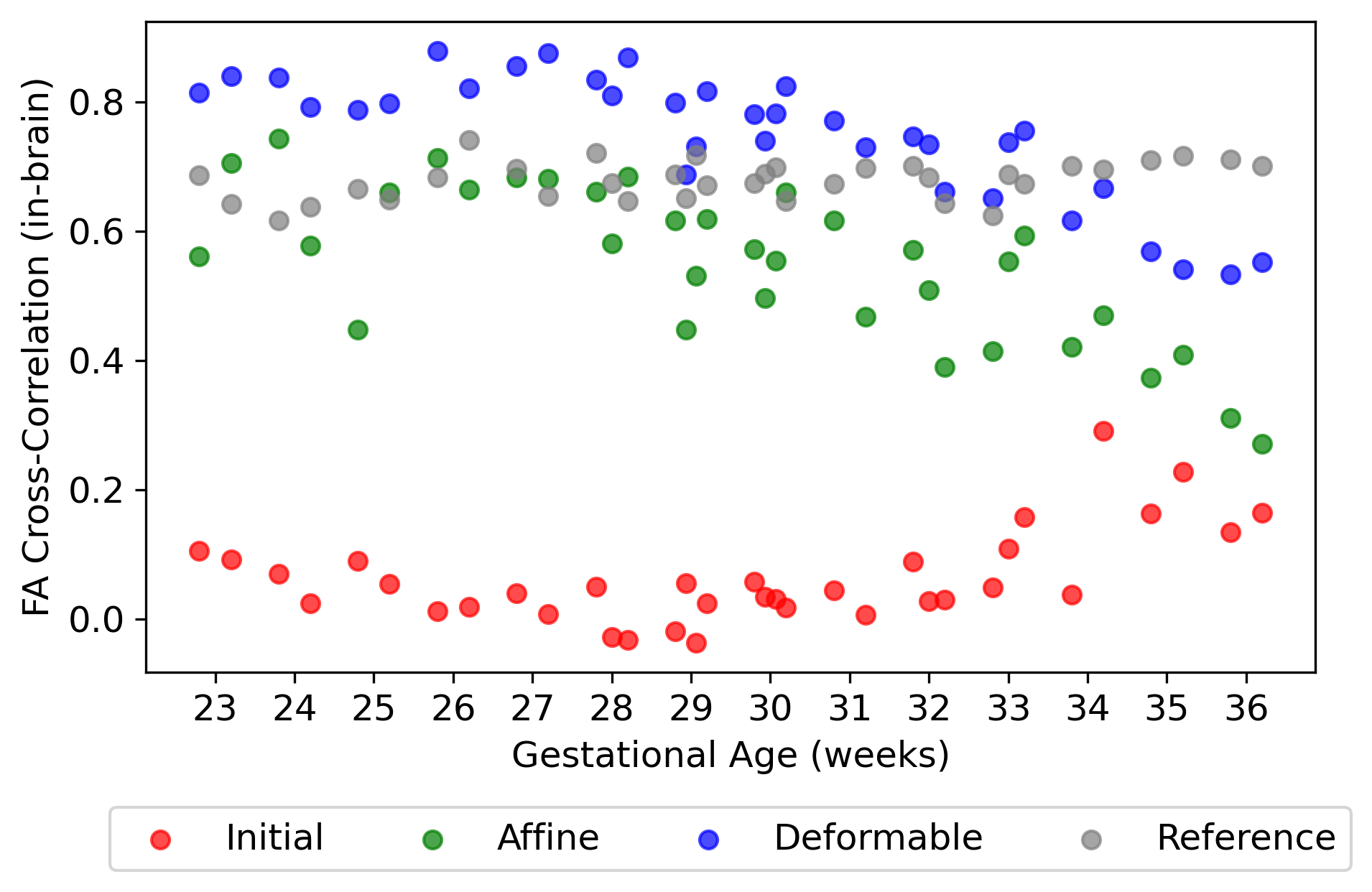}
    \caption{FA cross-correlation between the CRL fetal FA template and the dHCP FA map at different stages: initial, after affine registration, and after deformable registration using the proposed FetDTIAlign. The Reference value represents the cross-correlation between the dHCP FA map and itself after being translated by one voxel.}
    \label{fig:dhcp_cc}
\end{figure}

Before registration, the initial FA maps exhibited a very low cross-correlation (CC), in some cases reaching negative values (Figure~\ref{fig:dhcp_cc}). Across the entire test set, the initial CC was $0.0629\pm0.0708$, indicating poor alignment. Affine registration substantially improved the alignment, particularly for younger fetuses, whose smaller brain sizes initially led to less overlap with the template. After affine registration, the CC increased to $0.5491\pm0.1188$, demonstrating a statistically significant improvement in spatial correspondence ($p=1.8 \times 10^{-18}$). Deformable registration further refined the alignment, leading to a CC of $0.7493\pm0.0961$, indicating improved anatomical correspondence between the individual images and the template over the affine registration ($p=4.5\times 10^{-23}$). Notably, the reference similarity score, computed by artificially translating the images by one voxel as described above, was $CC=0.6787\pm0.0291$. Therefore, our deformable registration method achieved more accurate alignment than this minor misalignment ($p=6.6\times 10^{-4}$). It is also interesting to note that for affine registration the FA cross-correlation was higher on this unseen external dataset than on the internal BCH test dataset across all age groups. The same was true for deformable registration across all age groups except for 35 and 36 weeks.

\section{Discussion}

Reorienting vector images typically involves: i) Warping the image using high-quality interpolation methods such as trilinear or spline interpolation, ii) Computing the Jacobian of the displacement field to derive local rotation matrices, and iii) Applying these matrices to adjust the vector orientations within each voxel after interpolation. Although we considered improving reorientation by accounting for rotational effects during interpolation, practical implementation poses significant challenges to combine image warping and vector reorientation into a single operation. This difficulty arises because, while scalar image interpolation (e.g., intensity values) is straightforward, vector data requires careful handling to preserve orientation and magnitude during warping. Additionally, the spatial variation of the rotation applied to each voxel complicates the interpolation process further.

Similarly, while various data augmentation techniques exist, particularly for scalar images, we avoided applying unrealistic transformations, such as extreme shearing. Random intensity modifications and spatial deformations in diffusion tensor imaging can lead to physically implausible values. A valid diffusion tensor must be a positive definite matrix, and introducing random noise to its elements can disrupt this property. To maintain data integrity while enhancing model robustness, we restricted our augmentations to transformations that preserve the physical properties of the tensors, specifically translations and rotations, with careful correction of rotation effects after warping. Additionally, while generating synthetic images with poor SNR and low resolution could serve as a potential augmentation strategy, we chose not to apply such modifications. The fetal dMRI data used in this study already reflects the poor data quality encountered in practice, inherently exhibiting low SNR and resolution, which poses challenges for image preprocessing and registration approaches.

For high-quality FA images, direct registration using FA yielded acceptable results (Figure~\ref{fig:registration_fsl_tbss_t_manual_inhome}: Subject 4), with visually comparable outcomes across the methods. However, for FA images containing artifacts (Figure~\ref{fig:registration_dtitk_tbss_inhome_1.2mm}: Subject 2), diffusion tensor-based registration produced a more robust alignment of the overall brain shape. Despite this improvement, the internal structures remained misaligned, likely due to the nature of linear transformation and the missing details in the original image. Although VoxelMorph achieved very high similarity with the target image, as measured by cross-correlation, visual inspection of the registered images revealed several issues. In line with this finding, when excluding the diffusion tensor input (i.e., using only the FA and tract masks) we observed higher cross-correlation of FA but also the ``fake'' generated bright rims and ``white cracks'' on the registered images (Figure~\ref{fig:fa_nonlinear_GA25}). The addition of diffusion tensor decreased this metric, but led to smooth and reasonable deformation (Figure~\ref{fig:fa_nonlinear_GA25}, Figure~\ref{fig:skeleton_fa_nonlinear_GA25}).

The use of both FA images and the diffusion tensor, from which FA is derived, may seem redundant at first glance, given that FA is directly computed from the tensor. However, this dual-input approach did enhance the registration performance in our experiments. The diffusion tensor contains a more complete representation of the diffusion process as it includes the diffusivity values and the direction of the strongest diffusion. The FA image, on the other hand, provides a scalar summary of diffusion anisotropy, which simplifies the complexity of the tensor and highlights some of the key structural features such as major white matter tracts in a straightforward manner. This complementary relationship between the two inputs likely contributes to the improved performance, with the FA image facilitating robust overall alignment and the diffusion tensor map providing more detailed information. Thus, despite the apparent overlap, this combination leads to more accurate and reliable registration results by leveraging both the simplicity of FA and the detailed information in the tensor. In this context, other scalar indices such as mean diffusivity (MD) may offer additional benefit by robustly highlighting other key structures such as the ventricles or the cortical plate at different gestational ages. We leave an exploration of these other scalar maps to future works.

Interestingly, one could interpret the role of FA features as similar to an attention mechanism, as they emphasize structurally relevant regions that contribute to improved registration accuracy. A natural question that arises is whether an attention-based approach could serve as an alternative to FA in our framework. However, FA and attention mechanisms operate differently. FA, which encodes diffusion-based contrast, provides explicit structural landmarks. While self-attention mechanisms dynamically reweigh features based on learned dependencies, its ability to capture global structural information for registration remains uncertain. Investigating whether an attention-based approach could achieve comparable or superior performance to FA-based registration is an interesting direction for future research.

To ensure valid test results and prevent data leakage, for the internal (BCH) dataset we reserved three age groups (GA 25, 30, and 35 weeks) exclusively for testing. These groups represent distinct stages of brain development and varying tissue contrast. In fetal studies, a GA $\pm$ 1 or $\pm$ 2 week window is commonly used to account for biological variability, GA estimation accuracy, statistical stability, and imaging limitations. Consequently, fetuses were not confined to a single GA group, and the selected test groups collectively covered the full age spectrum (Figure~\ref{fig:training_test_split}).

Across the three age groups, most methods showed higher tract correspondence and FA similarity for younger age groups (Table~\ref{tab:registration_table_interSubjs}). This finding parallels a study investigating the reproducibility of tractography in the fetal brain, which reported a statistically significant decrease in reproducibility for fetuses in late gestational age ($>$30 weeks) compared to early gestational age ($\leq$ 30 weeks). \cite{xiao2024reproducibility}. This trend may be attributed to the fact that fetal brain initially exhibit simpler geometries, which gradually develop into more complex forms with greater individual variation. As a result, younger age groups are easier to align and reproduce, and present a higher achievable ceiling for performance. Additionally, the characteristics of the images across the age groups also play a role in determining the performance ceiling. In fetal brain development, FA values exhibit significant changes in different regions due to the maturation and structural organization of the brain's tissues over time. In earlier gestational ages (e.g., GA25), the simplicity of brain structures and the relatively higher FA in radial glial fibers make the images more homogeneous and straightforward for registration. As brain development progresses, the structures become more complex, and water diffusion in certain regions, such as the cortex,  becomes more isotropic \cite{mckinstry2002radial}. This results in lower FA values in those regions (Figure~\ref{fig:fa_affine_GA35}). As myelination progresses and dendritic branching becomes more prominent, the microstructural complexity increases, leading to more complex local FA patterns across the brain volume, making it harder to accurately align all these subtle local patterns.

In addition to the internal test set, we incorporated an external test set from the dHCP, ensuring a well-distributed selection across GA from 22 to 37 weeks. This external dataset serves as an independent validation set to assess the generalizability of the model across diverse acquisition conditions. Unlike the internal test set, subjects in the external test set were strictly assigned to their nominal gestational age. This ensures a direct GA-specific evaluation while minimizing potential confounding effects from age overlap. The inclusion of this independent dataset further demonstrates the robustness of our evaluation, providing insights into model performance across different imaging protocols and populations.

For the internal test set, we implemented a tract segmentation pipeline \cite{calixto2025detailed} to utilize the Dice score between registered tract masks as an evaluation metric for registration accuracy. The segmentation was carefully visually inspected by experts to ensure reliability. However, for the external test set, we opted to omit segmentation-based evaluation for two main reasons. First, optimizing a segmentation pipeline specifically for this dataset would be beyond the scope of this work. Ensuring adequate segmentation accuracy requires dataset-specific adaptations and verification of the results by human experts, which are not the focus of our study. Second, without thorough validation, the segmentation results would not be reliable enough to serve as an evaluation criterion. Segmentation uncertainties could introduce variability, confounding the Dice score and making it difficult to disentangle segmentation errors from registration errors. This would ultimately complicate the interpretation of registration performance.

Recent advancements in Transformer-based architectures for image registration have shown promise, particularly in capturing long-range dependencies and global context \cite{chen2022transmorph, mok2022affine}. Motivated by these developments, we investigated the use of a Transformer-based encoder for feature extraction in deformable registration, testing both concatenated input types and separate encoders for each input type, and different number of depth. However, our experiments did not reveal significant performance improvements over simple convolution-based encoders. One possible explanation is that convolutional networks are particularly adept at learning hierarchical spatial features, which are crucial for image registration tasks where local spatial consistency is key. While Transformers excel at capturing global relationships, the limited image quality of our fetal brain data, including low resolution, low SNR, and fewer detailed structures compared to the high-quality adult data typically used in Transformer-based applications, may have limited the benefits of using this architecture. Therefore, despite the theoretical advantages of Transformers, convolution-based encoders proved more effective for our specific registration task.

To encourage smooth deformations, penalizing negative Jacobian determinants could serve as an alternative smoothness loss term, or be combined with the spatial gradient term used in our loss function (Eq.~\ref{eq:Ldef}). However, since the percentage of negative Jacobian determinant (NDJ(\%)) is also used as an evaluation metric for all competing methods (Table~\ref{tab:registration_table_interSubjs}), explicitly optimizing the method on this criterion may introduce bias. Therefore, we opted to avoid direct penalization of NDJ\% to ensure a valid comparison on the deformation smoothness across methods.

A limitation of our evaluation was the use of DTI-TK both during the creation of the target template and during validation, inducing a positive bias for the method. However, DTI-TK represents a reliable tool for diffusion tensor image registration \cite{Pecheva2017TSA, Bach2014methodological}, and achieved the best results when we built the atlas. Therefore, using it as a baseline allows for a clearer demonstration of the performance of the proposed algorithm and its potential in future applications. For nonlinear registration methods, we used VoxelMorph to represent the performance of standard learning-based techniques for our tasks. While recent advances in learning-based medical image registration, such as transformer blocks for capturing long-range features and multi-scale iterative inference, show promise, these methods are primarily designed for high-quality single-channel gray-scale images and, like VoxelMorph, do not leverage orientation information in diffusion tensor. Additionally, VoxelMorph's auxiliary loss explicitly optimizes spatial correspondence between white matter tracts, integrating both global and regional information. Consequently, we believe VoxelMorph serves as a suitable reference for this study.

We have analyzed only DTI-derived measures, however, the framework allows data from other diffusion measures to be aligned to the group-specific template by applying the same transformations. Also, an interesting future direction is to apply the framework to patient datasets, enabling the investigation of group differences between normal and abnormal brain development at various gestational ages. By registering both patient and reference data to the same space, we can perform voxel-wise comparisons of diffusion measures. This will allow us to precisely quantify how early deviations from typical development manifest across gestational ages, contributing to a better understanding of abnormal brain development and potential early interventions.

\section{Conclusions}

In this study, we propose a novel deep learning approach, FetDTIAlign, for spatial normalization of fetal brain diffusion MRI, addressing the unique challenges posed by fetal imaging. While classical methods for diffusion MRI registration have been developed for adult brains, this work advances the state-of-the art for being the first to develop a deep learning framework specifically for fetal brain dMRI. The significantly lower image quality, distinct anatomy, and dynamic brain development require tailored solutions, different from those used for high-quality adult brain imaging or structural MRI. Our method was developed and validated across a wide range of gestational ages, from 23 to 36 weeks, covering a broad spectrum of early brain development that is not addressed in the literature. Furthermore, we included 60 white matter tracts in our analysis, offering a more extensive examination compared to studies that typically focus on specific tracts or age groups. Our results demonstrate the potential of the proposed deep learning method to improve spatial normalization in fetal brain dMRI. This method can lead to important future advancements in this area by enabling accurate and reliable cross-sectional and longitudinal studies.

\section*{Acknowledgements}

This research was supported in part by the National Institute of Neurological Disorders and Stroke under award number R01NS128281; the Eunice Kennedy Shriver National Institute of Child Health and Human Development under award number R01HD110772; National Institutes of Health grants R01LM013608, R01EB019483, R01NS124212, R01EB018988, R01NS106030, and R01EB032366, and the Office of the Director of the NIH under award number S10 OD025111. The content of this publication is solely the responsibility of the authors and does not necessarily represent the official views of the NIH.


\bibliographystyle{unsrt}
\bibliography{fetalTBSS_R2}






\end{document}